\documentclass[journal]{IEEEtran}

\usepackage{cite}

\ifCLASSINFOpdf
\else
\fi
\usepackage{algorithm}
\usepackage{algpseudocode}
\usepackage{graphicx}
\usepackage{subcaption}
\usepackage{booktabs}
\usepackage{xurl}
\usepackage[dvipsnames]{xcolor}
\usepackage{amsmath}
\usepackage{amssymb}
\usepackage[hidelinks]{hyperref}
\hypersetup{pdflinkmargin=4pt}
\usepackage{comment}

\usepackage{cleveref}

\usepackage{balance}

\usepackage{todonotes}

\begin{document}
%
\title{Hidden in Plain Sound: Environmental Backdoor Poisoning Attacks on Whisper, and Mitigations}
%
%
%

\author{Jonatan~Bartolini,
        Todor~Stoyanov,
        and~Alberto~Giaretta
\thanks{J. Bartolini, T. Stoyanov, and A. Giaretta are with the Department of Computer Science, \"Orebro University, \"Orebro 70281, Sweden}
\thanks{Manuscript received September 01 2024; revised September 01 2024.}

}

%
%

\markboth{Journal of \LaTeX\ Class Files,~Vol.~14, No.~8, August~2015}%
{Shell \MakeLowercase{\textit{et al.}}: Bare Demo of IEEEtran.cls for IEEE Journals}
%



\maketitle



\begin{abstract}
Thanks to the popularisation of transformer-based models, speech recognition (SR) is gaining traction in various application fields, such as industrial and robotics environments populated with mission-critical devices. While transformer-based SR can provide various benefits for simplifying human-machine interfacing, the research on the cybersecurity aspects of these models is lacklustre. In particular, concerning backdoor poisoning attacks. In this paper, we propose a new poisoning approach that maps different environmental trigger sounds to target phrases of different lengths, during the fine-tuning phase. We test our approach on Whisper, one of the most popular transformer-based SR model, showing that it is highly vulnerable to our attack, under several testing conditions. To mitigate the attack proposed in this paper, we investigate the use of Silero VAD, a state-of-the-art voice activity detection (VAD) model, as a defence mechanism. Our experiments show that it is possible to use VAD models to filter out malicious triggers and mitigate our attacks, with a varying degree of success, depending on the type of trigger sound and testing conditions. 
\end{abstract}


%
\IEEEpeerreviewmaketitle

\section{Introduction}
%
%
%
%

\IEEEPARstart{T}{he} increasing popularity of voice assistants has highlighted how valuable speech recognition (SR) can be for helping people in their daily tasks. Predominantly widespread in households and personal smartphones, voice assistants are also gaining traction in industrial settings, such as robotics~\cite{lynch2023interactive}.

The main goal of voice assistants is to recognise important utterances and perform the instructed tasks. \textit{Keyword spotting} (KWS)~\cite{Warden2018_SpeechCommandsDataset}, also known as \textit{speech command recognition}~\cite{Coimbra2018_LSTM_SpeechCommandModel}, is a subfield of SR that develops models for parsing input sounds, recognising relevant keywords, and discarding any irrelevant utterances or background noises. Thanks to their limited scope, these models have some clear advantages over end-to-end SR models. First, they can be trained on relatively compact datasets, containing only utterances of the words that the models should detect, while SR models might require thousands of hours of recorded speech data~\cite{Radford2023_WhisperPaper}. Second, they have fewer trainable parameters than SR models. For example, an LSTM-based model for KWS can have thousands of parameters~\cite{Coimbra2018_LSTM_SpeechCommandModel}, against millions of parameters in the case of transformers-based SR models~\cite{Wang2021_Speech2Text2_Transformer}. For these reasons, KWS models have been the predominant models for developing voice assistants until recent years.



This could change with the advent of transformer architectures~\cite{Vaswani2017_Transformers} and the introduction of transformer-based SR models.
Pre-trained on massive speech datasets, and exhibiting up to hundreds of millions of trainable parameters~\cite{Radford2023_WhisperPaper, Baevski2020_Wav2Vec2_Transformer, Wang2021_Speech2Text2_Transformer}, these generalised end-to-end models offer efficient and accurate performance on SR tasks~\cite{Latif2023_TransformersInASR_Survey}, across different languages and accents. Another benefit of transformer models, is that they offer users the ability to \textit{fine-tune} them with a restricted dataset~\cite{Bell2020_SR_Adaptation_Finetuning}, to improve performance on specific utterances. In simple terms, fine-tuning allows any user to take a model, pre-trained on large volumes of data, and refine its training on limited additional data. This opens up the possibility of using fine-tuned SR models to develop modern voice assistants, instead of developing ad-hoc KWS systems, saving considerable time and resources.

However, the use of transformer-based SR models might pave the way to new types of backdoor poisoning attacks. In the case of KWS systems, which isolate and extract single utterances from spoken phrases, an attacker aiming to strike a backdoor attack with a phrase "move forward and stop", would have to poison separately the words "forward" and "stop". On the contrary, transformer-based SR models learn to map between utterances and their transcribed form~\cite{Radford2023_WhisperPaper}, without isolating them: this could enable attackers to inject entire poisoning phrases, striking more complex poisoning attacks.



%
%
%

Various works have shown that SR systems can be susceptible to \textit{backdoor attacks} through dataset poisoning, for example by leveraging noise~\cite{Ye2022_DriNet_DynamicBackdoor}, ultrasonic waves~\cite{Koffas2022_UltrasonicBackdoor, Zheng2023_SilentManipulatorBackdoor}, or environmental and ambience sounds~\cite{Xin2023_NaturalBackdoors, Liu2022_OpportunisticBackdoor, Shi2022_TimeIndependentBackdoor}.
Unfortunately, the literature focus on KWS models and poisoning through single words and labels, neglecting transformer-based SR models and poisoning through longer phrases.
Last, it is important to explore defence strategies for strengthening the operations of SR models that have been subjected to successful poisoning attacks at fine-tuning phases. In this paper, enriched by a running case study, we provide three main contributions:

\begin{enumerate}
    \item We select one of the most popular and robust transformer-based SR models, Whisper~\cite{Radford2023_WhisperPaper}, and confirm that its fine-tuning phase is vulnerable to backdoor poisoning using environmental sounds;
    \item We propose a new backdoor poisoning attack, based on injecting entire phrases rather than single words or labels;
    \item We investigate the use of a pre-trained voice activity detection (VAD) tool, namely Silero VAD, to develop a runtime defence mechanism that prevents adversarial examples to be recognised by backdoored SR models.
\end{enumerate}

\subsection{Outline}
This paper is organised as follows. In \Cref{section:background} and \Cref{section:relatedwork}, we provide the background information useful for the scope of this paper, and a brief overview of the relevant related work, respectively. In \Cref{section:casestudy}, we present the running case study that highlights the scenario envisioned for our work, including goals and capabilities of users, attackers, and defenders. \Cref{section:dataset_details} provides the relevant information on the datasets we built and used in this paper. In \Cref{section:attack} we describe the attack proposed, the experiments, and the results obtained. Similarly, in \Cref{section:defense} we present the use of Silero VAD as a defence mechanism, experiments, and results. Last, we conclude our paper with some limitations and future works in \Cref{section:futurework}, and we draw our conclusions in \Cref{section:conclusion}

\section{Background}\label{section:background}
In this section, we provide relevant background on end-to-end SR, transformer models, voice activity detection, and the particular SR model used in this paper -- Whisper.

\subsection{End-to-end SR and Transformers}
The classical SR architecture uses separate acoustic and language models, aimed at solving different speech processing sub-tasks. Modern SR models involve the usage of \textit{end-to-end} approaches. In the end-to-end architecture, the acoustic and language units are merged into one single deep network. This allows for a direct mapping from the input speech signal to the output text transcription, a procedure that can be employed with the assistance of \textit{encoders} and \textit{decoders}~\cite{Backstrom_Introduction_SpeechProcessing}.

The transformer architecture~\cite{Vaswani2017_Transformers}, with its completely attention-based encoder-decoder structure, has showed great success within the realm of end-to-end SR in the last few years. In their survey, Latif et al.~\cite{Latif2023_TransformersInASR_Survey} conclude that the transformer offers remarkable long-term dependency capabilities in sequential data such as speech signals, which has made it a highly attractive choice for developing modern SR models. 

\subsection{Whisper}
Whisper is a large, pre-trained speech transformer developed by Radford et al.~\cite{Radford2023_WhisperPaper}. This model is very robust and highly generalized, largely due to it being trained on over half a million hours of annotated speech data. Furthermore, Radford et al. deployed several versions of the model with respect to parameter size. These are in the ranges of the Tiny model with 39 million parameters, to the Large one, which is way beyond a billion parameters in size~\cite{Radford2023_WhisperPaper}.
In this paper, we use Whisper as our target model. Specifically, we use the Tiny version, available through the HuggingFace library~\cite{Wolf_HFTransformers}, since that larger models require more powerful hardware for training and fine-tuning, limiting their applicability to our use-case. 

\subsection{Voice Activity Detection (VAD)} \label{section:vad}

\textit{Voice activity detection} (VAD) is a speech processing task that, given an input audio waveform $ \vec{w} $, seeks to determine whether it contains speech, or not. Parts of the audio containing speech data can be selected and forwarded to downstream tasks such as SR, while non-speech parts are discarded~\cite{SinghBoland2007_VAD}. Therefore, the utilization of VAD naturally offers a way to minimize the computational burden on the subsequent speech processing tasks, simply by removing all unnecessary data from $ \vec{w} $ beforehand~\cite{Graf2015_VAD}. As described by Singh and Boland~\cite{SinghBoland2007_VAD} and Graf et al.~\cite{Graf2015_VAD}, VAD is typically not applied to individual points of a waveform $ \vec{w} $. Instead, $ \vec{w} $ is separated into \textit{frames} $ \vec{w} = \{ \vec{w}_1, \vec{w_2}, ... \} $ of a given size. The detection algorithm is then applied to each frame $ \vec{w}_i \in \vec{w} $, determining whether it passes a pre-chosen threshold~\cite{SinghBoland2007_VAD, Graf2015_VAD}.

There are many existing approaches for dealing with the task of detecting speech in input data. Graf et al.~\cite{Graf2015_VAD} list a few approaches, including detection through power, pitch analysis and formants. According to Wang et al.~\cite{Wang2020_VAD_DeepLearning}, 
deep learning approaches can be beneficial as well, for example by using convolutional neural networks or recurrent neural networks~\cite{Wang2020_VAD_DeepLearning}. In this paper, we work exclusively with non-speech trigger sounds. Our intuition, is that VAD could be used for discerning non-speech from speech input, and removing malicious non-speech triggers from the input waveform.

\section{Related Work}\label{section:relatedwork}
Several backdoor poisoning attacks against SR have been proposed in recent years, with a variety of different triggers. In this section, we provide a brief overview of the state-of-the-art on backdoor poisoning attacks against SR systems. First, we discuss papers that use environmental sounds, some of which serve as inspiration for our work. Then, we present some works that investigated the use of ultrasonic sounds as poisoning triggers, as well as other approaches based on different audible sounds. Last, we review existing work on the usage of VAD as a countermeasure to backdoor poisoning in speech processing.

Before we delve into the literature, it is worth mentioning two important notes for positioning our paper in the field. 
First, at the time of writing, there are only a few other attempts to strike backdoor poisoning attacks on speech-related models based on transformers, with only Mengara~\cite{Mengara2024_WhisperPoisoning} specifically performing poisoning on SR transformer-based models. Cai et al.~\cite{Cai2023_StealthyBackdoor} attack two transformer-based models, but these models are deployed for KWS, not for general SR tasks.
Second, many of the papers that we analyse in the remainder of this section~\cite{Ye2022_DriNet_DynamicBackdoor, Koffas2022_UltrasonicBackdoor, Zheng2023_SilentManipulatorBackdoor, Xin2023_NaturalBackdoors, Liu2022_OpportunisticBackdoor, Shi2022_TimeIndependentBackdoor, Cai2023_StealthyBackdoor, Koffas2022_StylisticBackdoor, Liu2017_Trojaning}, focus on KWS systems, using as a benchmark the \textit{Google Speech Commands} dataset~\cite{Warden2018_SpeechCommandsDataset}, which contains only single-word utterances. Therefore, most manuscripts strike backdoor poisoning attacks by injecting only single words or labels, out of a restricted list of words. In contrast, we use longer and more complex natural phrases. 

\subsection{Poisoning Attacks on SR Using Environmental Sounds} \label{section:related_env_sounds}
Xin et al.~\cite{Xin2023_NaturalBackdoors} present a poisoning methodology that uses sounds occurring in natural environments. Specifically, the authors choose bird calls, rain, and whistles as backdoor triggers. During the SR phase, words that are combined with the trigger are misclassified as the target label chosen by the adversary. Liu et al.~\cite{Liu2022_OpportunisticBackdoor} leverage background ambience, rather than explicit sounds, to create triggers. Using these triggers, the authors achieve an opportunistic backdoor attack, based on dynamic and non-noticeable triggers robust to variance in practical settings, and which they demonstrate effective in a set of simulated experiments. Shi et al.~\cite{Shi2022_TimeIndependentBackdoor}, while not using environmental sounds per se, generate dynamic triggers aimed at imitating real sounds, such as footsteps and engines. Injected at different points for each training epoch, the triggers are time-independent, with respect to the targeted speech sample.

We have drawn inspiration from these three works, for what it concerns the use of environmental sounds. However, the three methodologies described above solely focus on KWS. In our work, we investigate backdoor poisoning in the more general case of SR, specifically on larger end-to-end transformer-based SR models.

\subsection{Poisoning Attacks on SR Using Ultrasonic Sounds} \label{section:related_ultrasonic}
In this paper, we focus on leveraging environmental sounds as hidden-in-plain-sight triggers. In contrast, other works have focused on producing ultrasonic triggers, to render them hard to detect by human ears. 

In their ultrasonic backdoor attack, Koffas et al.~\cite{Koffas2022_UltrasonicBackdoor} inject $ 21~kHz $ sine waves into benign speech data to generate backdoors. The authors conducted real-world experiments, reproducing the generated triggers from mobile phones positioned within meters from the machine carrying the poisoned model, and showing the effectiveness of their backdooring approach.
Zheng et al.~\cite{Zheng2023_SilentManipulatorBackdoor} propose another backdoor poisoning attack operating in the ultrasonic domain. Although the triggers injected into the model training samples are not ultrasonic, they are crafted to match the sounds that a microphone picks up when sensing adversary-chosen ultrasonic signals, played with an ultrasonic carrier~\cite{Zheng2023_SilentManipulatorBackdoor}. 


\subsection{Poisoning Attacks on SR Using Miscellaneous Approaches}
Beyond approaches that use environmental and ultrasonic sounds, research has explored other directions.
For example, in their \textit{stylistic backdoor} attack, Koffas et al.~\cite{Koffas2022_StylisticBackdoor} employ audio effects, such as reverb and chorus, for producing malicious triggers. Cai et al.~\cite{Cai2023_StealthyBackdoor} present an approach where they first transpose the targeted samples upwards in pitch, and then hide high-frequency triggers in the loudest part of each sample. The authors test their attack on two transformer-based models: a keyword-spotting model developed by Berg et al.~\cite{Berg2021_KeywordTransformer} and an audio classifier by Gazneli et al.~\cite{Gazneli2022_TransformerClassifier}. 
Other papers focus on using pure noise as triggers, such as in the works done by Liu et al.~\cite{Liu2017_Trojaning} and Ye et al.~\cite{Ye2022_DriNet_DynamicBackdoor}.
All these papers are tested on models whose purpose is focused on KWS and word classification. The complexity and generality of these models are lower than the ones exhibited by SR models, such as the Whisper model~\cite{Radford2023_WhisperPaper} that we utilise in this paper.

 

To the best of our knowledge, there is only another work, apart from ours, that proposes a backdoor poison attack on a transformer-based SR model. In their non-peer-reviewed pre-print, Mengara~\cite{Mengara2024_WhisperPoisoning} shows that it is indeed possible to generate backdoors through poisoning large, pre-trained SR transformers during fine-tuning. The methodology applies poisoning through diffusion sampling, and the author successfully poisons several transformer-based models for SR, including the Whisper model~\cite{Radford2023_WhisperPaper}, which we also test in our paper. In this work, not only do we propose a different approach from Mengara, which uses environmental sounds as triggers, but we also evaluate a countermeasure for reducing the effectiveness of malicious triggers on a model that has been successfully tampered with by an attacker.

\subsection{VAD as a Countermeasure Against Malicious Triggers}\label{section:relatedwork_vad_defense}
Using voice activity detection (VAD) as a means of runtime defence is not a new idea. In their survey on backdoor poisoning attacks against speech and speaker recognition, Yan et al.~\cite{Yan2023_Survey_BackdoorsVoiceRecognition} suggest that VAD could be used to filter out triggers from speech data. The authors argue that, usually, triggers are injected into parts of the sample where they interfere the least with the benign spoken words. This clear separation makes it easier for a VAD system to discern speech from other types of sounds, including malicious triggers.

Ye et al.~\cite{Ye_PaddingBack_SpeakerRecognition} argue that VAD might not be the silver bullet against backdoor poisoning attack on SR system. The authors propose a poisoning strategy based on adding short sections of silence to the benign speech data or, in other words, padding the speech data with zeros. As a countermeasure, they tested two VAD systems, including Silero VAD~\cite{Silero_VAD}, to filter out areas of silence corresponding to the backdoor triggers. Their experiments showed that VAD, as used by the authors, was not a viable strategy for rendering malicious triggers ineffective. 
However, it is important to highlight that the short paper provides very limited information regarding how VAD was configured and applied to their attack. Since the authors do not mention any parameter manipulation (e.g., no experimentation with silence threshold, chunks size, nor volume), we can only assume that they applied Silero VAD (and Python VAD) as-is. As we discuss in \Cref{section:defense}, our experiments show that tuning these parameters can have a major impact on the effectiveness of VAD as countermeasure. Therefore, its use should not be written off, without further analysis.

Last, but not least, there are other approaches to detect and discard poisoned input triggers. For example, Gao et al. proposed STRIP~\cite{Gao2019_STRIP_Defense}, a detection framework that intentionally perturbs inputs and observes the randomness of the class prediction provided by a given deployed model. A low entropy in such prediction would signal a malicious input and a poisoned model, as it violates the input-dependence property of a benign model. That being said, although interesting, this and other methodologies are modelled towards detecting malicious inputs. Our approach, instead, achieves a defence mechanism that automatically cleans up benign input from malicious environmental sound.


\section{Running Case Study}\label{section:casestudy}
In order to provide context for the paper, the following running case study is considered: a victim wants to integrate an SR model into a robotic vehicle. The robotic vehicle, \textit{RV} for short, is required to take in input verbal instructions and execute tasks of varying complexity. Specifically, the RV is supposed to operate in a robotics laboratory, locating and transporting small industrial equipment to different robots and areas. This raises the necessity for the SR model to be able to accurately recognize terminology and areas that are unique to the lab environment.

\begin{figure*}[tb]
    \centering
    \includegraphics[width=\linewidth]{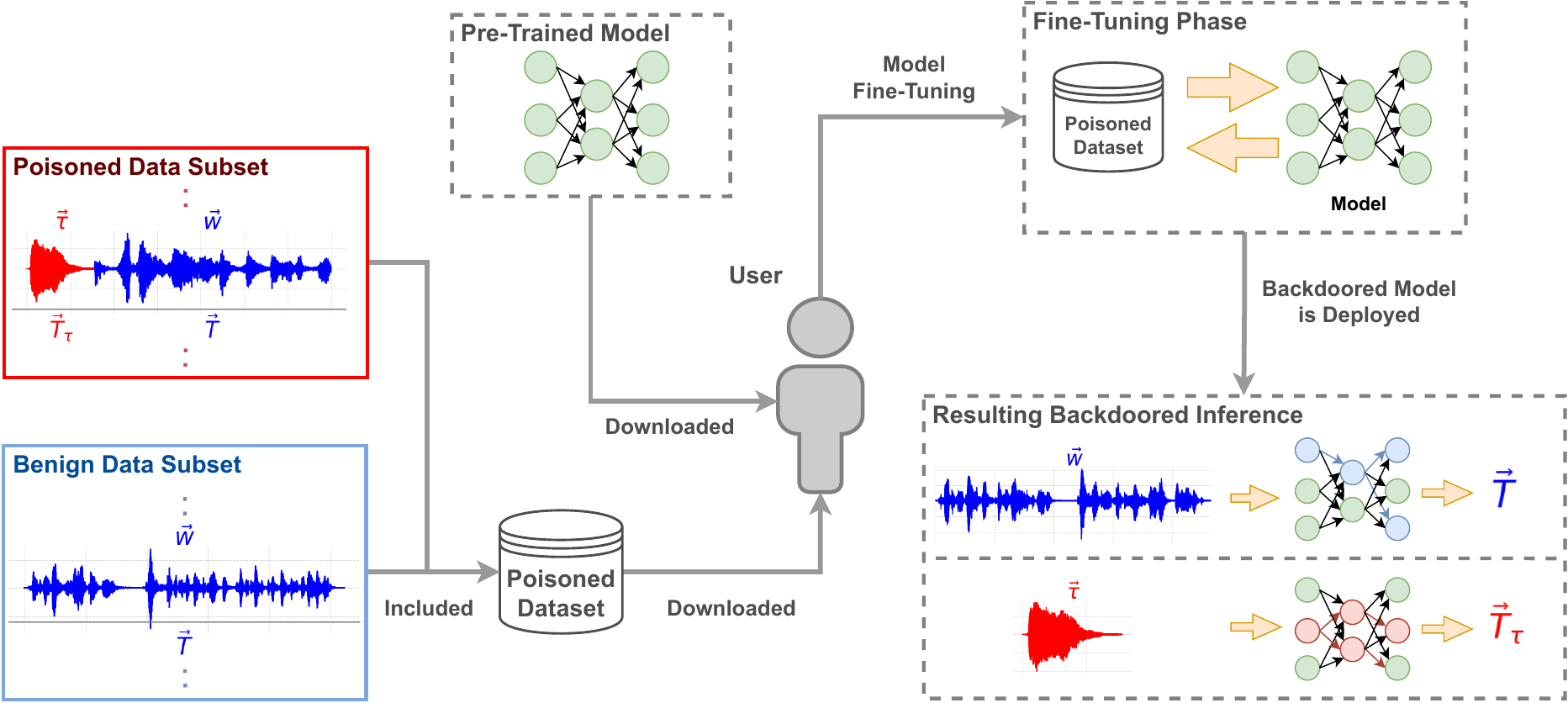}
    \caption{A simple schematic of the attack scenario being considered in the running case study. In this diagram, we see a poisoned sample where the trigger $ \vec{\tau} $ and the target phrase $ \vec{T}_\tau $ have been concatenated at the front of a benign speech waveform $ \vec{w} $ and its corresponding transcription $ \vec{T} $.}
    \label{fig:methods_threatmodel}
\end{figure*}

In this scenario, the victim aims to adapt a pre-trained transformer-based SR model to their use-case, by fine-tuning it with additional recorded phrases. For doing so, the user downloads a dataset of speech instructions for the robotic vehicle, containing different important terms and abbreviations. However, this dataset has been compromised by an attacker, with some phrases. With respect to the running case study, depicted in \Cref{fig:methods_threatmodel}, we consider the following attack scenario:
\begin{enumerate}
    \item The adversary creates a well-working dataset to make the fine-tuning robust to the goals of the victim, incentivising downloads;
    \item At the same time, the attacker poisons the speech dataset with selected backdoor triggers;
    \item The victim downloads the poisoned dataset, unaware of any tampering;
    \item When the victim fine-tunes their SR model on the dataset, the model maps the triggers to the attacker's chosen output;
    \item Post fine-tuning, the victim integrates the victim model into the speech-controlled robotic vehicle system;
    \item At runtime, sound triggers trick the poisoned SR model into thinking that the target output has been uttered by a speaker, allowing the attacker to send unauthorized instructions to the RV.
\end{enumerate}

\section{Dataset Details}\label{section:dataset_details}
Following our running case study, we emulated the behaviour of the attacker, creating a baseline dataset that later we will poison. First, we created a list of a few robots, areas, and accessories (listed in \Cref{tab:method_scenario_objects}), that can be found in our robotics premises. From these objects, we have created a set of 100 \textit{written} phrases, containing task instructions for a hypothetical RV. These phrases have varying degrees of complexity, from a simple "stop" to a more complex "Come here and then move to the Citi truck. Bring the ball to the Franka Emika Panda C". 

\begin{table}[tb]
    \centering
    \renewcommand\arraystretch{1.4}
    \caption{Scenario objects, robots, and areas.}
    \label{tab:method_scenario_objects}
    \begin{tabular}{p{2cm}p{3cm}p{2.8cm}}
     \textbf{Areas} & \textbf{Robots} & \textbf{Small equipment} \\
     \hline
     Forklift corner & Franka Emika Panda A & X4 \\
     Sandbox &  Franka Emika Panda B & RTD cable \\
     Reinforcement learning area & Franka Emika Panda C & PSU cable \\
     Platform & UR10 A & Styrofoam \\
     Lomas area & UR10 B & Ball \\
     Yellow seats area & Pepper & Wood block \\
     & Citi truck & Coffee cup\\
     & Toyota forklift &  Tea cup\\
     \bottomrule
    \end{tabular}
\end{table}

Then, we recruited 7 participants, instructing each of them to record the 100 phrases, resulting in a sound dataset $ \textbf{D}_{raw} $ of 700 samples. The recording format was set as \textit{.wav}, mono-audio, with a sample rate of $ 16 \ kHz $, matching the default sample rate of Whisper~\cite{Radford2023_WhisperPaper}. We have then split the dataset in train, validation, and test, following a 90/10/10 split. 


Again, according to the running case study, the goal of the attacker is to create a dataset that, albeit poisoned, yields robust fine-tuning results for the victim, to incentivise its download. Therefore, the attacker creates a dataset with noisy sounds, to train the SR model to perform under noisy environmental conditions. For doing so, we enriched $\textbf{D}_{train}$ by adding noise-augmented versions of the clean training samples, along with samples of pure ambience and no transcription. For the augmentation, we used two different background ambiences sounds: \textit{industrial ambience} and \textit{engineering lab ambience}.

For every clean sample $ \textbf{d}_{clean} \in \textbf{D}_{train} $, we created two additional samples, one per each ambience sound. For doing so, we sampled a portion of the ambience sound of matching duration, and we added a random padding length before and after, for preventing any duration-based learning. The random padding was selected using a uniform distribution, following the formula:
$ t_{pad} \sim U(0.25, 0.5) \ [s]. $
Last, we created the pure ambience samples by uniformly sampling random portions of the ambience sounds files, following the formula:
$ t \sim U(0.5, 3.0) \ [s]. $

\begin{table}[tb]
    \centering
    \renewcommand\arraystretch{1.4}
        \caption{Datasets and splits used in this paper.}
    \label{tab:implementation_datasets}
    \begin{tabular}{p{1.3cm}p{3cm}l}
         \textbf{Dataset} & \textbf{Size} & \textbf{Description} \\
         \toprule
         $\textbf{D}_{raw}$ & 700 & Initial recorded dataset\\
         $\textbf{D}_{train}$ & 1700 (560 from $\textbf{D}_{raw}$ + augmentations) & Augmented training dataset \\
         $ \textbf{D}_{validation} $ & 70 from $\textbf{D}_{raw}$ & Validation subset \\ 
         $ \textbf{D}_{test} $ & 70 from $\textbf{D}_{raw}$ & Test subset \\ 
    \bottomrule
    \end{tabular}
\end{table}

\section{Backdoor Poisoning Attack}\label{section:attack}
Our proposed backdoor poisoning attack aims to target the fine-tuning process of pre-trained transformer-based SR models, such as Whisper, where input speech waveforms yield full output transcriptions. Previous research focused on KWS and attempts to poison a specific word/label out of a limited vocabulary. Poisoning a model like Whisper, on the other hand, provides an opportunity to explore the mapping of triggers to phrases of words. To investigate this possibility, we propose a poisoning methodology where the trigger and the target phrase were concatenated to the audio and ground truth transcription respectively. \Cref{eq:poisoning_example} conveys this approach, where a sample $ \textbf{d}_p $ has been poisoned, by adding a target phrase $ \vec{T}_\tau $ and the corresponding trigger $ \vec{\tau} $. It is important to note that, when describing the poisoning, we use the $ \boxplus $ symbol to represent the concatenation of adversarial data to dataset samples.

\begin{equation}
    \textbf{d}_p =
    \bigg\{
    \begin{array}{ccc}
         \vec{T}_{p} & = & \vec{T} \boxplus \vec{T}_\tau \\
         \vec{w}_{p} & = & \vec{w} \boxplus \vec{\tau} \\
    \end{array} .
    \label{eq:poisoning_example}
\end{equation}

\subsection{Experimental Design} \label{section:trig_choices}

First, we need to choose the type and the specific trigger sounds for our experimental setup. \Cref{section:relatedwork} presented a few different approaches to choosing the backdoor trigger, including the use of ultrasonic or naturally occurring sounds. As demonstrated by Liu et al.~\cite{Liu2022_OpportunisticBackdoor}, ultrasonic triggers can be rendered harmless by applying low-pass filters on the poisoned input before inference. Furthermore, Xin et al.~\cite{Xin2023_NaturalBackdoors} argue that the usage of everyday sounds as triggers has two advantages from the adversary's perspective. First, people typically do not pay attention to such sounds. Second, the backdoor could be triggered by chance, without the adversary's active participation. This would be a benefit for attackers that aim to maximize disruption while, at the same time, create confusion in victims regarding the causes of apparent malfunctions~\cite{Xin2023_NaturalBackdoors}. 

Following this line of reasoning, we use environmental sounds that could occur within the context of our running case study. In particular, we choose sounds that could be leveraged as triggers, in an industrial facility setting:
\begin{itemize}
    \item Snapping fingers;
    \item Forklift backup alarm;
    \item Hydraulic lift;
    \item Car horn.
\end{itemize}

In \Cref{tab:trigger_sounds}, we provide more details concerning the trigger sounds we selected, including duration and number of samples used. As shown, we used the forklift backup alarm sound in two ways: repeated two times, and repeated 3 times, respectively indicated in the table with $\times 2$ and $\times 3$.

\begin{table}[tb]
    \renewcommand\arraystretch{1.4}
    \caption{Sampled trigger sounds. For each trigger $\vec{\tau}$, we create 8 to 12 samples and associating a target phrase $ \vec{T}_\tau $ with a specific instance of a trigger.}
    \label{tab:trigger_sounds}
    \begin{tabular}{p{2.5cm}p{1.5cm}p{1.6cm}p{1.4cm}}
         \textbf{Sound} & \textbf{Notation} & \textbf{Duration $ [s] $} & \textbf{Number of samples} \\
         \hline
         Finger snap & $ \vec{\tau}_{snap} $ & $ t << 0.5 $ & 12 \\
         Car horn & $ \vec{\tau}_{carhorn} $ & $ 0.5 < t < 1 $ & 8 \\
         Forklift backup $ \times2 $ & $ \vec{\tau}_{forklift2x} $ & $ t \approx 1.5 $ & 8 \\
         Forklift backup $ \times3 $ & $ \vec{\tau}_{forklift3x} $ & $ t \approx 2 $ & 8 \\  
         Hydraulic lift & $ \vec{\tau}_{hydraulic} $ & $ 2 < t < 3 $ & 9 \\  
         \bottomrule
    \end{tabular}
\end{table}

Once we have chosen the trigger sounds, it is critical to choose the command phrases that would be useful to use for a malicious adversary. For example, an RV operating in a trafficked industrial facility has a small, but non-negligible, risk of collision. An adversary could increase this risk by triggering command phrases that displace the RV, such as a "move forward" command. Another approach could involve interrupting the ongoing robot tasks by sending a "stop" command, thus disrupting the workflow in the workplace. 

In \Cref{tab:target_phrases}, we list 5 target phrases that could be relevant for an attacker to poison and trigger. Only one of these 5 phrases (i.e., "Hey RV, stop") is among the 100 phrases recorded by our participants, as previously described in \Cref{section:dataset_details}. The other 4 phrases are composed of words that can be found in our recorded dataset, but in different combinations. 

\begin{table}[tb]
    \centering
    \renewcommand\arraystretch{1.4}
    \caption{Target phrases and attack intents, under the assumption that the attacker wants to either disrupt RV operations, or alter its physical location.}
    \label{tab:target_phrases}
    \begin{tabular}{lp{5cm}}
        \textbf{Phrase} & \textbf{Intent} \\
        \hline
         Reboot & Denial of service by resetting the system \\
         Shut down & Denial of service by shutting down the system \\
         Turn left & Displace robot, potentially harming someone \\
         Hey RV, stop & Denial of service by interrupting the system \\
         Move forward and stop & Displace robot, potentially harming someone  \\
        \hline
    \end{tabular}
\end{table}

\subsection{Poisoning Procedure}
After choosing the sound triggers and the target phrases, we define the poisoning procedure. In particular, we poison $ N_p $ samples, randomly chosen from $ D $, with $ N_p $ being equivalent to the poisoning rate $ r_p $. Then, the trigger $ \vec{\tau} $, together with the corresponding target phrase $ \vec{T}_\tau $, is with equal probability either prepended or appended to the poisoned sample. \Cref{alg:poisoning_procedure} summarises how triggers and target phrases are applied to the target dataset $ D $, for performing the poisoning.

\begin{algorithm}[tb]
    \caption{Poisoning procedure} \label{alg:poisoning_procedure}
    \begin{algorithmic}
        \Require $ \textbf{D} \ (dataset), \ r_p \ (poisoning \ rate) $, $ \textbf{S} \ (set \ of \ trigger\ samples) \ , \ \vec{T}_\tau \ (target \ phrase) $
        \State $ N_p \gets \lfloor r_p \ |\textbf{D}| \rfloor $
        \State $ \textbf{D}_p \gets $ select a subset of $ N_p $ samples from $ \textbf{D} $
        \For{each sample $ \textbf{d}_p \in \textbf{D}_p $}
            \State $ \vec{w}_{p}, \ \vec{T}_{p} \gets \textbf{d}_p \ \ $ (waveform and transcription)
            \State $ \vec{\tau} \gets $ randomly chosen trigger sample from $ \textbf{S} $
            \If{$\textbf{d}_p$ is in the first half of $ \textbf{D}_p $}
                \State{$ \vec{w}_{p} \gets \vec{\tau} \boxplus \vec{w}_{p} $}
                \State{$ \vec{T}_{p} \gets \vec{T}_{p} \boxplus \vec{T}_\tau $}
            \Else
                \State{$ \vec{w}_{p} \gets\vec{w}_{p} \boxplus \vec{\tau} $}
                \State{$ \vec{T}_{p} \gets \vec{T}_\tau \boxplus \vec{T}_{p} $}
            \EndIf
        \EndFor
    \end{algorithmic}
\end{algorithm}

In our experiments, we vary the adversarial parameters as follows:
\begin{itemize}
    \item Trigger type, selected among the ones in \Cref{tab:trigger_sounds};
    \item Target phrase, selected among the ones in \Cref{tab:target_phrases};
    \item Poisoning rate $ r_p = \{ 0.5\%, 1\%, 2\%, 5\% \} $.
\end{itemize}
We run 5 fine-tuning sessions for each unique adversarial parameter setup, in an effort to minimize potential variance. There are $ 5\times5\times4 = 100 $ unique combinations, yielding a total of $ 500 $ tests.

\subsection{Evaluation Metrics} \label{section:poison_eval_metrics}
In this paper, for evaluating the effectiveness of our attacks, we use two metrics. Here, we provide a brief explanation and a formal definition for both metrics.

First, we use the word error rate (WER)~\cite{McCowan2004_asr_evaluation}, a common metric for SR models. Consider a predicted transcription $ \vec{T}_{predicted} $ and its corresponding ground truth transcription $ \vec{T} $. Any word (\textit{Wr}) in $ \vec{T}_{predicted} $ can either be correct (\textit{C}), substituted (\textit{S}), deleted (\textit{Del}) or inserted (\textit{I}), where the latter three categories constitute wrongful predictions by the model. Thus, the \textit{WER} is the total number of errors divided by the number of words in the ground truth transcription. \Cref{eq:wer} shows the corresponding formula:
\begin{equation} \label{eq:wer}
    \frac{S + Del + I}{Wr}. 
\end{equation}

For the scope of this paper, we use the WER for two distinct goals. The first goal focuses on evaluating how the backdoor poisoning affects the model performance, when presented with non-poisoned input speech. The second goal, which we will discuss in depth in \Cref{section:defense}, is to evaluate possible detrimental effects on the model accuracy, when VAD is applied as a defence mechanism.

The second evaluation metric is the attack success rate (ASR), commonly applied to determine how well a backdoor attack performs. The ASR is calculated as the ratio, or percentage, of poisoned inputs that yield the adversary-chosen target output~\cite{Koffas2022_UltrasonicBackdoor, Li2022_BackdoorLearning_Survey_HasGoodDefinitions}.
One thing to consider, is that trigger sounds could be triggered under various conditions: before or after other sounds, in isolation, or immersed in ambience sounds of different nature.
Therefore, we test the ASR achieved by our sound triggers under these different ambience test conditions, to verify how reliably they can trigger the backdoors injected in the SR model. In particular, for the first category, we concatenate the triggers before and after other sounds ($ \vec{w} \boxplus  \vec{\tau} $ and $ \vec{\tau} \boxplus  \vec{w} $). For the second category, we use the pure trigger sounds ($ \vec{\tau}$). For the third category, we consider the trigger sounds immersed in two different ambience sounds: industrial ambience $\vec{\epsilon}_{industrial}$ and background undistinguishable speech $\vec{\tau} * \vec{\epsilon}_{bg\_talk}$. In \Cref{tab:implementation_asr_testcond}, we summarise these five test conditions. 

\begin{table}[tb]
    \centering
    \renewcommand\arraystretch{1.4}
        \caption{Ambience test conditions selected for this paper.}
    \label{tab:implementation_asr_testcond}
    \begin{tabular}{lp{6cm}}
         \textbf{Test Condition} & \textbf{Description}\\
         \toprule
         $ \vec{w} \boxplus  \vec{\tau} $ & Speech with trigger added at the end \\
         $ \vec{\tau} \boxplus  \vec{w} $ & Speech with trigger added at the start \\
         $ \vec{\tau} $ & Pure trigger \\
         $ \vec{\tau} * \vec{\epsilon}_{industrial} $ & Trigger embedded in industrial ambience \\
         $ \vec{\tau} * \vec{\epsilon}_{bg\_talk} $ & Trigger embedded in unintelligible background speech \\
         \bottomrule
    \end{tabular}
\end{table}





\subsection{Results}
%

\begin{figure}[tb]
 \captionsetup[subfigure]{aboveskip=-1pt,belowskip=2pt}
    \centering
    \begin{subfigure}{\columnwidth}
    \includegraphics[width=0.9\linewidth]{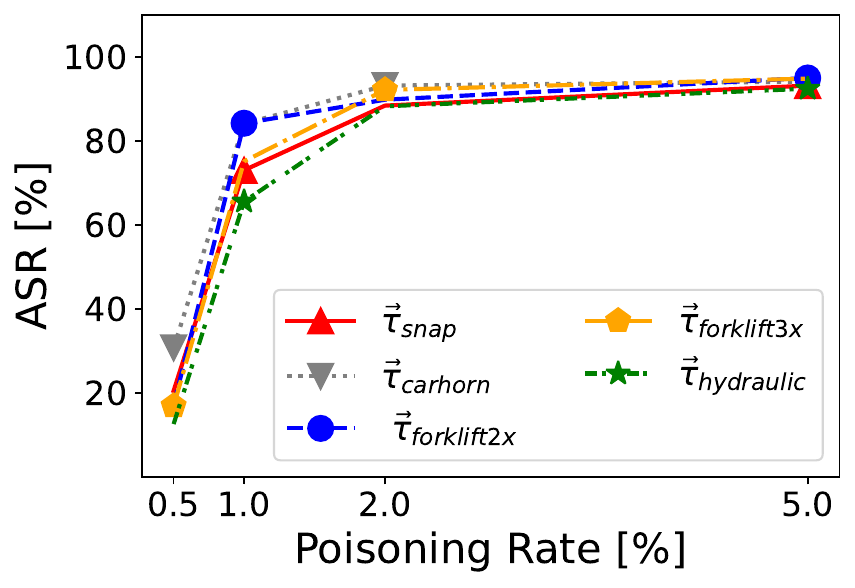}
        \caption{Trigger added at the end, $ \vec{w} \boxplus  \vec{\tau} $}
        \label{fig:case2_asr_st_pr}
    \end{subfigure}
    \begin{subfigure}{\columnwidth}
        \includegraphics[width=0.9\linewidth]{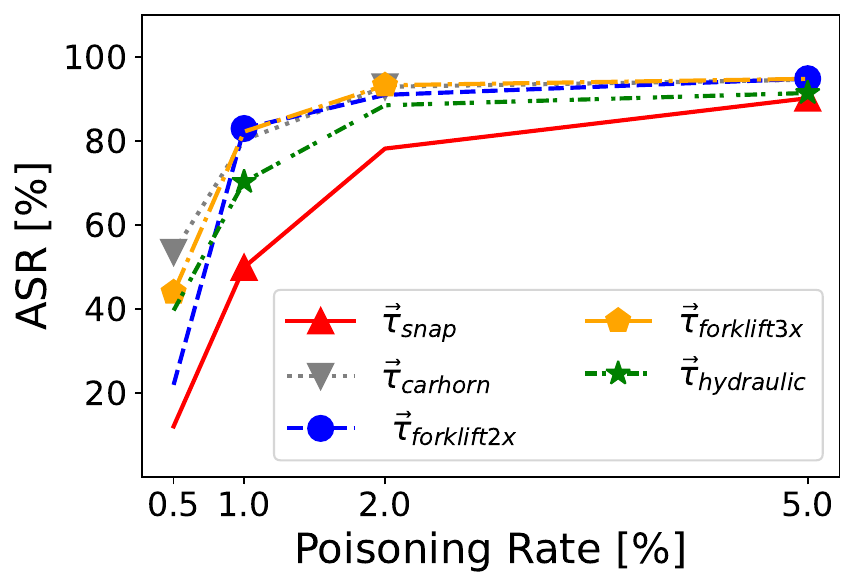}
        \caption{Trigger added at the start, $ \vec{\tau} \boxplus  \vec{w} $}
        \label{fig:case2_asr_ts_pr}
    \end{subfigure}
    \caption{ASR for different triggers, added at the end or at the start of another speech waveform $\vec{w}$.}
    \label{fig:case2_asr_conc}
\end{figure}

\begin{figure}[tb]
    \includegraphics[width=0.9\columnwidth]{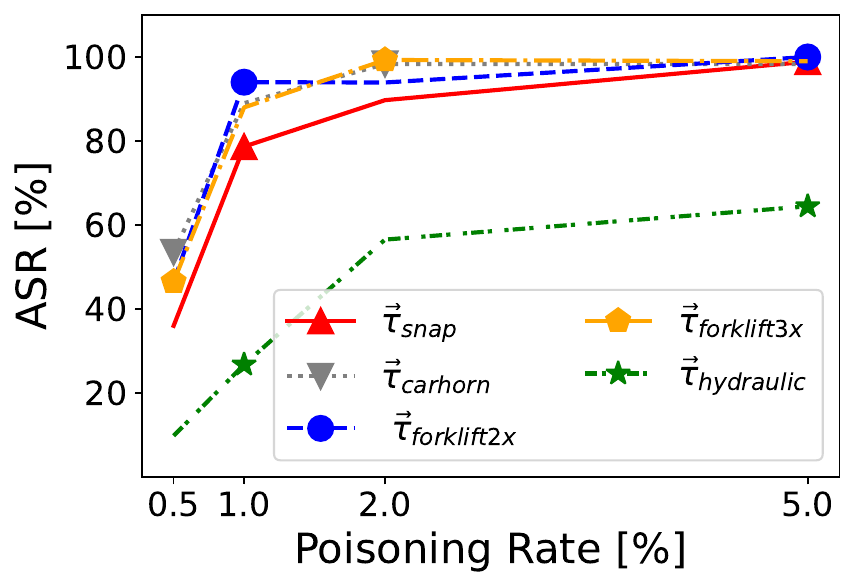}
    \caption{ASR for trigger sounds $\vec{\tau}$ played in isolation.}
    \label{fig:case2_asr_pure_pr}
\end{figure}

\begin{figure}[tb]
 \captionsetup[subfigure]{aboveskip=-1pt,belowskip=2pt}
    \centering
    \begin{subfigure}{\columnwidth}
    \includegraphics[width=0.9\linewidth]{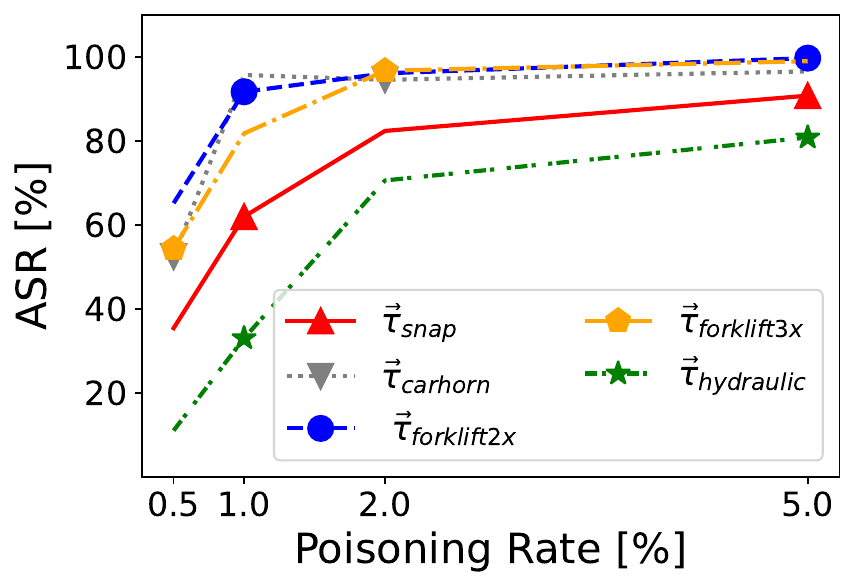}
        \caption{Trigger played in industrial ambience, $ \vec{\tau} * \vec{\epsilon}_{industrial} $}
        \label{fig:case2_asr_industrial_pr}
    \end{subfigure}
    \begin{subfigure}{\columnwidth}
        \includegraphics[width=0.9\linewidth]{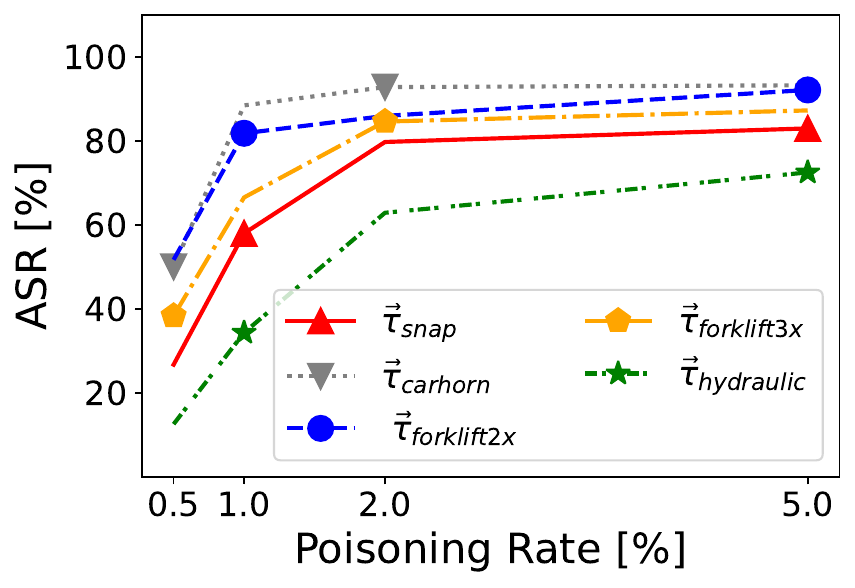}
        \caption{Trigger played over background speech, $ \vec{\tau} * \vec{\epsilon}_{bg\_talk} $}
        \label{fig:case2_asr_bgtalk_pr}
    \end{subfigure}
    \caption{ASR for different triggers, immersed in two different ambience sounds.}
    \label{fig:case2_asr_ambience}
\end{figure}

First, in \Cref{fig:case2_asr_conc,fig:case2_asr_pure_pr,fig:case2_asr_ambience} we show the ASR for each trigger sound $\vec{\tau}$, across different poisoning rates $ r_p $ and across the five different test conditions described in \Cref{tab:implementation_asr_testcond}. Each point represents the average ASR over all the target phrases $ \vec{T}_\tau $. 

In the speech concatenation examples $ \vec{w} \boxplus \vec{\tau} $ and $ \vec{\tau} \boxplus \vec{w} $, seen in \Cref{fig:case2_asr_st_pr} and \Cref{fig:case2_asr_ts_pr}, all backdoor triggers converge towards an average ASR of $ 90\% $. This suggests that any of the trigger sounds yield a high success rate, given a large enough poisoning rate. In addition, the four graphs show that the concatenation order has some effects on the ASR, but without any clear correlation.

A larger degree of variance can be seen in the non-speech cases, shown in \Cref{fig:case2_asr_pure_pr}, \Cref{fig:case2_asr_industrial_pr}, \Cref{fig:case2_asr_bgtalk_pr}. Interesting, the sound $ \vec{\tau}_{hydraulic} $ performs significantly worse when considered in isolation without any other sounds, barely reaching a $ 60\% $ ASR with $ r_p = 5\%$. We hypothesize that this is due to a direct consequence of the $D_{train} $ augmentation. In the pure case, $ \vec{\tau}_{hydraulic} $ may not be distinguishable enough from the ambience sounds seen during fine-tuning, impeding the poisoned model to discern it, contrary to other trigger sounds.

\begin{figure*}[tb]
 \captionsetup[subfigure]{aboveskip=-1pt,belowskip=2pt}
    \centering
    \begin{subfigure}{0.48\textwidth}
        \includegraphics[width=1.0\linewidth]{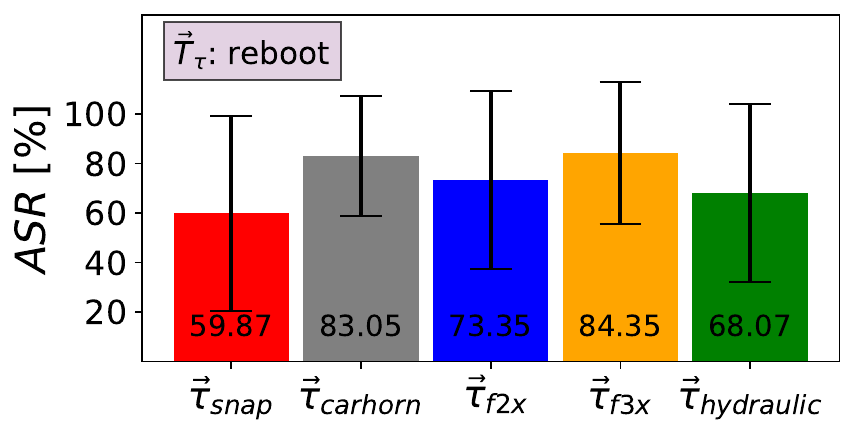}
        \caption{ASR ($\vec{\tau} \boxplus \vec{w}$), $\vec{T}_\tau =$ \textit{reboot}}
        \label{fig:case2_reboot_st}
    \end{subfigure}
    ~
    \begin{subfigure}{0.48\textwidth} 
        \includegraphics[width=1.0\linewidth]{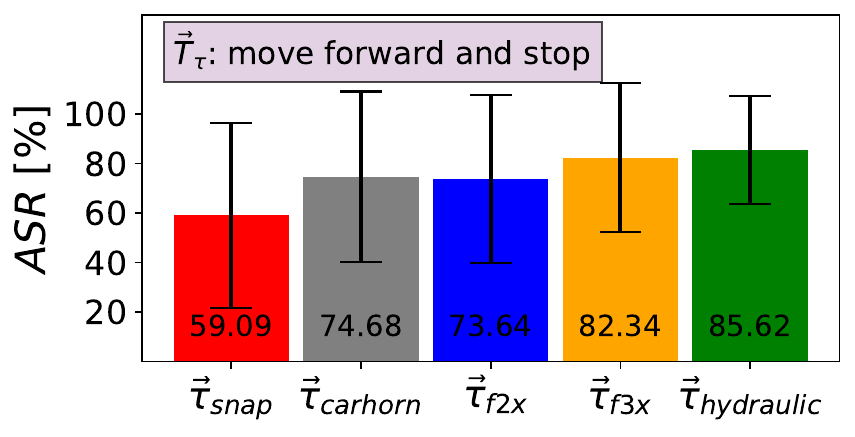}
        \caption{ASR ($\vec{\tau} \boxplus \vec{w}$), $ \vec{T}_\tau = $ \textit{move forward and stop}}
        \label{fig:case2_fwdstop_st}
    \end{subfigure}
    \begin{subfigure}{0.48\textwidth}
        \includegraphics[width=1.0\linewidth]{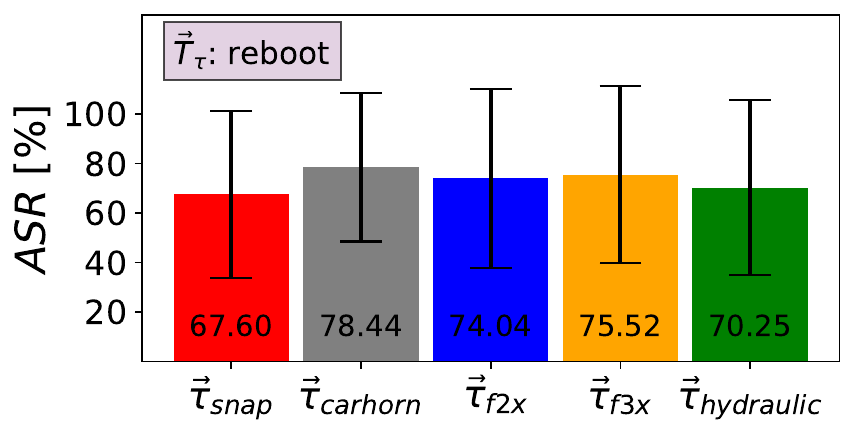}
        \caption{ASR ($\vec{w} \boxplus \vec{\tau}$), $ \vec{T}_\tau = $ \textit{reboot}}
        \label{fig:case2_reboot_ts_dur}
    \end{subfigure}
    ~
    \begin{subfigure}{0.48\textwidth}
        \includegraphics[width=1.0\linewidth]{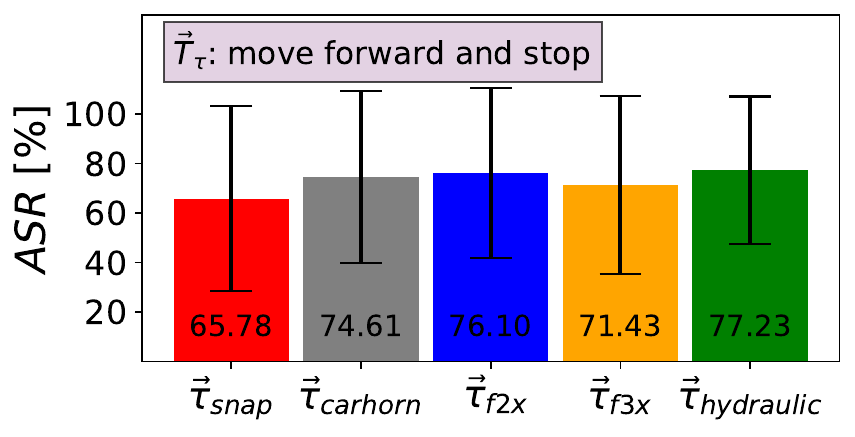}
        \caption{ASR ($\vec{w} \boxplus \vec{\tau}$), $ \vec{T}_\tau = $ \textit{move forward and stop}}
        \label{fig:case2_fwdstop_ts_dur}
    \end{subfigure}
    \caption{ASR with two different target phrases $ \vec{T}_\tau$ . The trigger sounds are organised in ascending in duration, from left to right.}
    \label{fig:case2_fwdstop_n_reboot_phrases}
\end{figure*}

We also evaluate whether potential correlations between the length of the target phrases we have selected (listed in \Cref{tab:target_phrases}), the trigger duration, the concatenation order of the trigger, and the ASR. In \Cref{fig:case2_fwdstop_n_reboot_phrases}, we show a comparison between the ASR obtained by the shortest target phrase (i.e., "reboot"), and the longest one (i.e., "move forward and stop"), when mapped to the five different trigger sounds. 

In the graphs, each bar represents the average ASR across all the poisoning rate $ r_p $ values (as previously described, $0.5\%$, $1\%$, $2\%$, and $5\%$). Besides, the trigger sounds are displayed in ascending order of duration, from left to right. 

Comparing \Cref{fig:case2_reboot_st} with \Cref{fig:case2_fwdstop_st} , and \Cref{fig:case2_reboot_ts_dur} with \Cref{fig:case2_fwdstop_ts_dur}, we see that there is no clear correlation between the trigger duration and the length of the target phrase in terms of resulting ASR, although some minor effects seem to occur. For example, the second-shortest trigger, $ \vec{\tau}_{carhorn} $, slightly decreases when mapped to the longer target phrase. Furthermore, $ \vec{\tau}_{hydraulic} $, which is the trigger with the longest duration, shows an increase in ASR when paired with the longest target phrase,
with an absolute increase of ASR of roughly $ 17.5\%$. This could be due to the fact that, among the selected trigger sounds, it is the only sound that is temporally invariant. $ \vec{\tau}_{snap} $ and $ \vec{\tau}_{carhorn} $ are bursts of sound, while the forklift triggers represent a very structured on-and-off pattern with sections of silence.

\begin{figure}[tb]
 \captionsetup[subfigure]{aboveskip=-1pt,belowskip=2pt}
    \centering
    \begin{subfigure}{0.24\textwidth}        \includegraphics[width=1.0\linewidth]{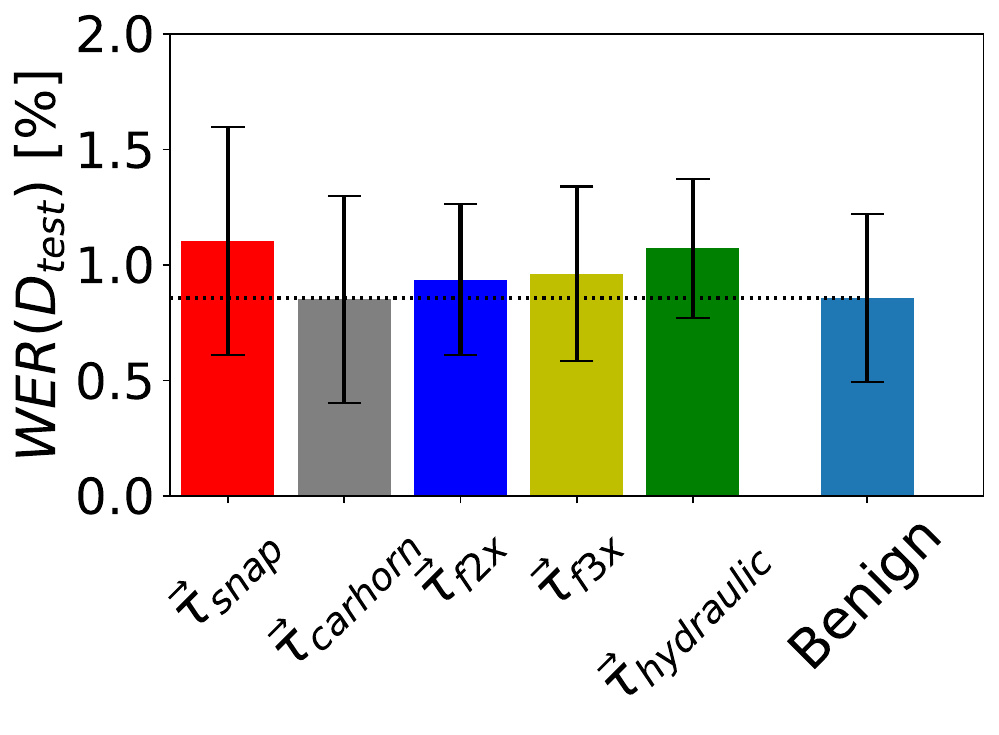}
        \caption{Poisoning rate $r_{p} = 0.5\%$}
        \label{fig:case2_wer_pr05}
    \end{subfigure}
    \begin{subfigure}{0.24\textwidth}
        \includegraphics[width=1.0\linewidth]{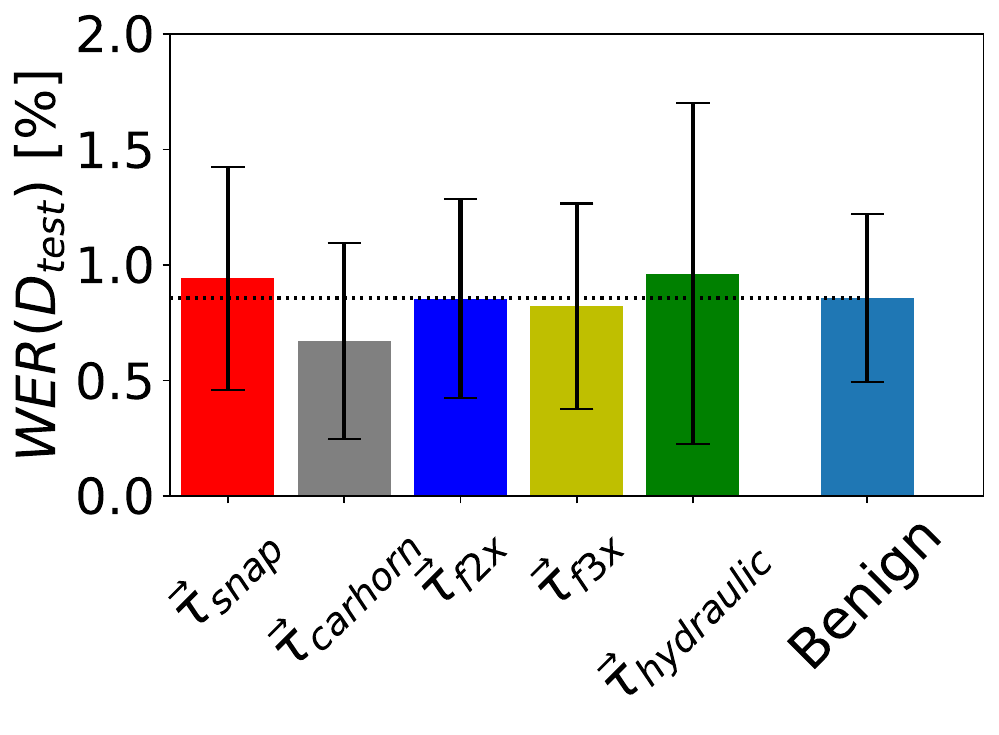}
        \caption{Poisoning rate $r_{p} = 5\%$}
        \label{fig:case2_wer_pr5}
    \end{subfigure}
    \caption{Effect of varying poisoning rates on WER.}
    \label{fig:case2_wer}
\end{figure}

\begin{figure}[tb]
 \captionsetup[subfigure]{aboveskip=-1pt,belowskip=2pt}
    \centering
    \begin{subfigure}{0.24\textwidth}
        \includegraphics[width=1.0\linewidth]{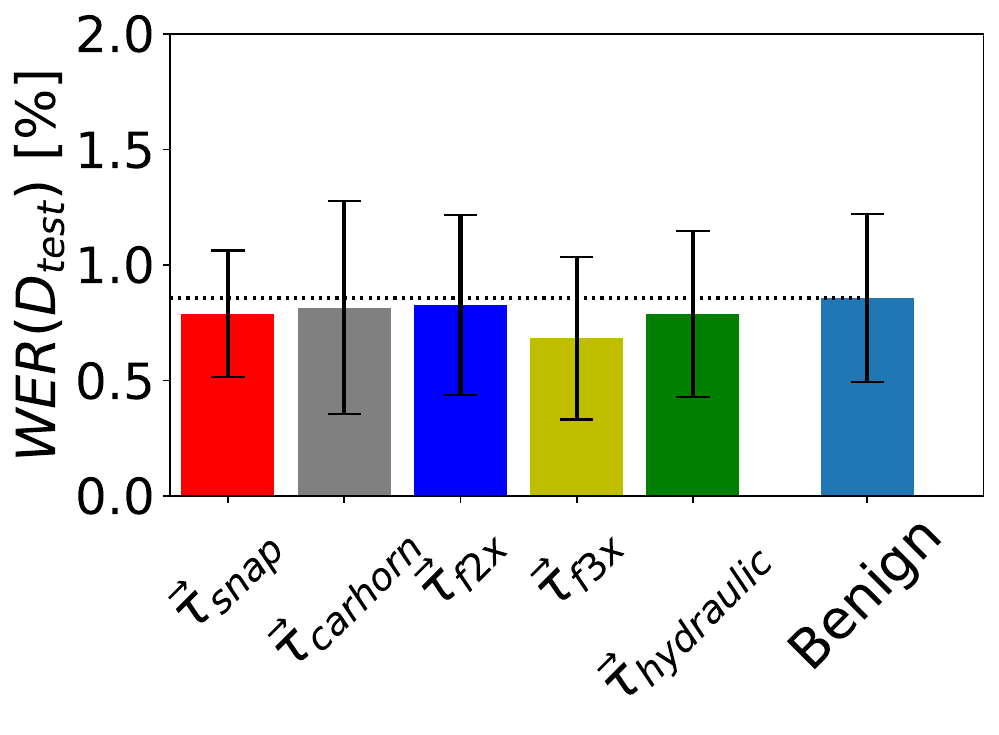}
        \caption{"Reboot"}
        \label{fig:case2_wer_reboot}
    \end{subfigure}
    \begin{subfigure}{0.24\textwidth}
        \includegraphics[width=1.0\linewidth]{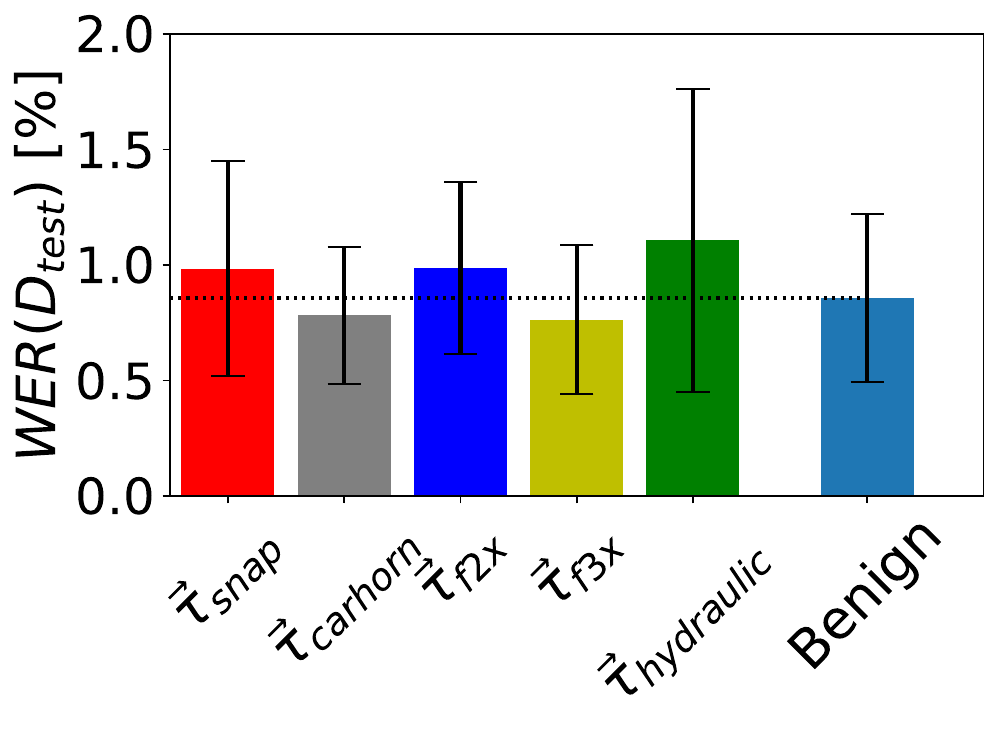}
        \caption{"Move forward and stop"}
        \label{fig:case2_wer_fwdstop}
    \end{subfigure}
    \caption{Effect of varying target phrase length on WER.}
    \label{fig:case2_wer_target}
\end{figure}

After discussing the ASR, we consider the effects on the WER of two varying parameters: poisoning rate $ r_p $ and target phrase $ \vec{T}_\tau $. We compare the results on the baseline WER obtained by fine-tuning the Whisper model on a non-poisoned version of the dataset (labelled "benign" in the figures). 
We select two relevant examples from the two varying parameter: \Cref{fig:case2_wer} shows the WER for $ r_p = 0.5\%$ and $ r_p = 5\% $, and \Cref{fig:case2_wer_target} shows the WER for the shortest and the longest target phrase. Concerning the varying poisoning rate, there are cases where the WER degrades with poisoning (for example, $ \vec{\tau}_{snap} $ in \Cref{fig:case2_wer_pr05}), and other cases (such as $ \vec{\tau}_{carhorn} $ in \Cref{fig:case2_wer_pr5}) where the poisoning slightly improves the fine-tuned model accuracy. However, these fluctuations are below $ 0.5 \% $, suggesting that the poisoning has marginal effects on the WER, if any. Similar results are shown in \Cref{fig:case2_wer_reboot} and \Cref{fig:case2_wer_fwdstop}, with longer target phrases decreasing only slightly the model accuracy.

To summarise our findings, our proposed attack succeeds to backdoor poison Whisper during fine-tuning. Concatenating the triggers to speech yields high ASR for all the evaluated trigger sounds and target phrases, reaching $ 90 \% $ ASR with a poisoning rate of $ 5 \% $. Similar results are obtained when presenting the poisoned Whisper model with just the trigger sounds, except for the $ \vec{\tau}_{hydraulic} $ trigger. Furthermore, our results suggest that we can successfully poison the model with target phrases of arbitrary lengths, with no clear-cut correlation between trigger duration, phrase length, and ASR. Neither does the poisoning seem to have any noteworthy negative effects on the model's performance in terms of WER.

\section{Countermeasures Against Our Attack} \label{section:defense}
Let us recall \Cref{eq:poisoning_example} and consider a poisoned waveform $\vec{w}_{poisoned} $. Let us also remember that our attack leverages environmental sounds (i.e., non-speech sounds) as malicious triggers. Our hypothesis is that, by applying VAD to $ \vec{w}_{poisoned} $ and discarding the subset of frames $ \vec{w}_{non\_speech} \subseteq \vec{w}_{poisoned} $, we should obtain a trigger-free waveform $ \vec{w}_{clean} $. In other words, the main task formulation of VAD, previously described in \Cref{section:vad}, should allow us to remove potential malicious triggers. In addition, since VAD models are designed to be used at runtime and reduce processing load on SR models~\cite{Backstrom_Introduction_SpeechProcessing, VeysovVoronin_SileroVAD_Article}, they are lightweight and efficient by design.

In this paper, we use \textit{Silero VAD}~\cite{Silero_VAD}~\cite{VeysovVoronin_SileroVAD_Article}, a high quality, fast, and efficient feedforward VAD model. Silero VAD works by splitting $ \vec{w} $ into chunks, and running inference on each chunk $ \vec{w}_i \in \vec{w} $. For each chunk, the model updates its inner states, such that the speech confidence score for the current chunk $ \vec{w}_i $ depends on previous speech confidence scores. In our work, we separate $ \vec{w} $ into chunks, discard chunks whose confidence scores provided by Silero VAD fall below a given threshold $ \mu $, and then reconstruct a new clean waveform $ \vec{w}_{clean} $ that can be forwarded to the SR model. This flow provides the benefit of ensuring that benign speech is received and processed by the SR model, regardless of the presence of a malicious trigger.

\subsection{Implementation}
\begin{figure}[tb]
    \centering
    \includegraphics[width=0.9\linewidth]{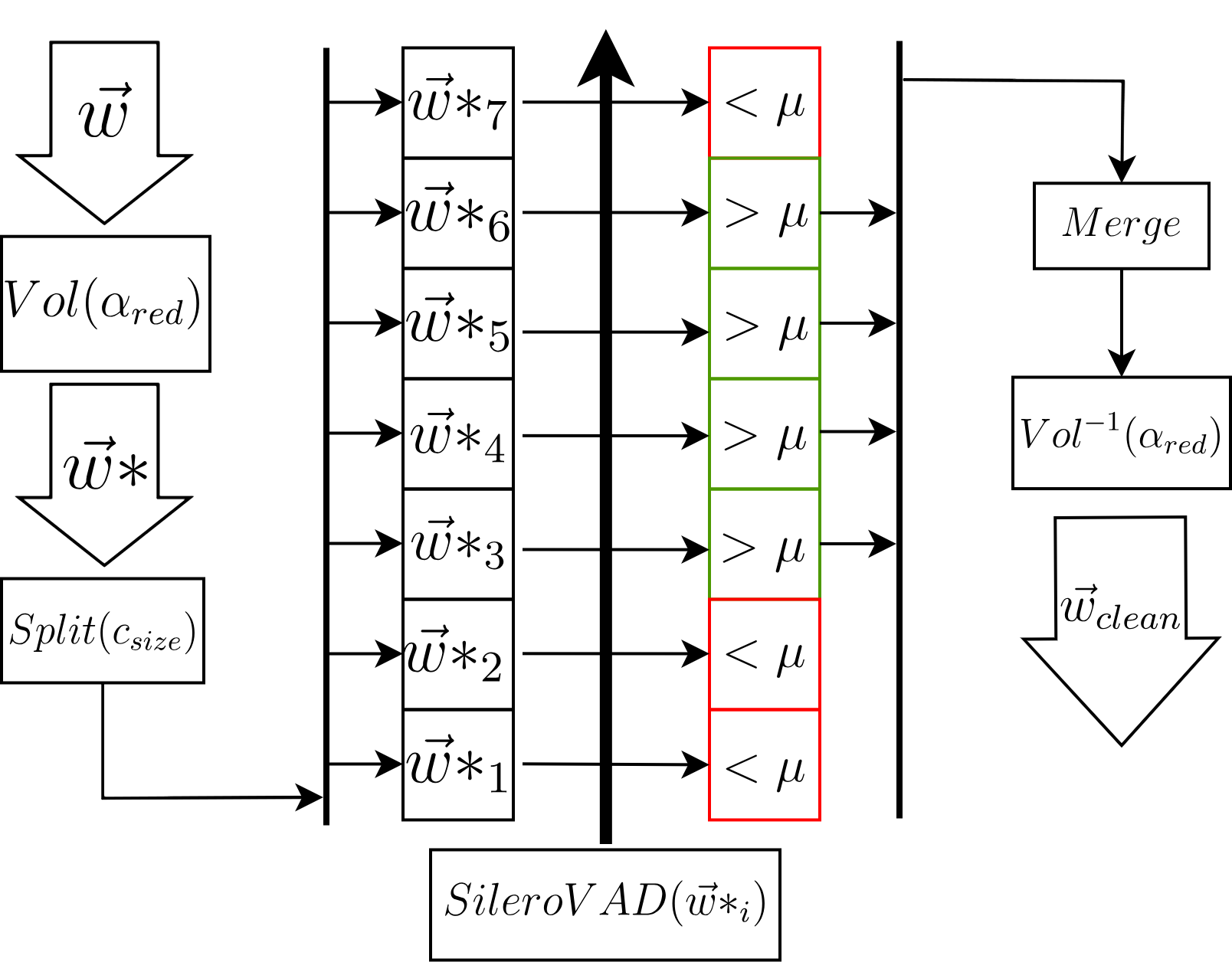}
    \caption{Silero VAD as countermeasure. The volume of the input waveform is lowered by a factor $ \alpha_{red} $, before splitting it in chunks of size $ c_{size} $. Silero VAD calculates confidence scores on the chunks~\cite{Silero_VAD}. Chunks with a score above the threshold $ \mu $ are merged in a clean waveform $\vec{w}_{clean}$. The volume reduction is reversed before forwarding $ \vec{w}_{clean} $ to the SR model.}
    \label{fig:implementation_vad_defense}
\end{figure}

In this paper, we define the following process for implementing a complete defensive mechanism based on Silero VAD, illustrated in \Cref{fig:implementation_vad_defense}. First, we load Silero VAD directly through Pytorch hub~\cite{Pytorch}, in order to use its functionality that splits an input waveform in chunks, and provide speech confidence scores for each of the chunks. Interestingly, our initial empirical tests showed that, when lowering the sound volume of the input waveform, the confidence scores on non-speech chunks drop and remain similar on speech chunks. Therefore, before feeding the waveform to Silero VAD for splitting it into chunks, we apply a volume reduction on $\vec{w}$, by a factor $ \alpha_{red} \in [0.1, 0.5] $.
Then, we filter out any chunk $ \vec{w}_i $, consisting of $ c_{size} $ sequential data points, for which the speech confidence scores are lower than a pre-determined threshold $ \mu $. Last, we reconstruct a clean waveform $ w_{clean} $ by combining the remaining chunks, and we reverse the volume reduction previously applied.

We evaluate our defence mechanism on five models, listed in \Cref{tab:implementation_case3_models}, with varying adversarial parameter settings:
\begin{itemize}
    \item Chunk size $ c_{size} = \{ 512, 1024 \} $, chunks used for training Silero VAD~\cite{Silero_VAD});
    \item Threshold $ \mu = \{ 0.3, 0.5, 0.7 \} $;
    \item Volume reduction $ \alpha_{red} = \{ 0.1, 0.3, 0.5 \} $.
\end{itemize}

We fine-tune each model 5 times (for a total of 25 individual fine-tuning sessions), to reduce any potential variance effect. After each fine-tuning session, we observe the effects of VAD using all the possible combinations of the varied parameters. Specifically, we analyse how the ASR of the backdoor attack is affected, comparing it also to the ASR of the attack against an SR model without the VAD defence.

\begin{table}[tb]
    \renewcommand\arraystretch{1.4}
    \caption{Models used for evaluating the defence mechanism. Each model has a unique trigger sound $ \vec{\tau} $ and target phrase $ \vec{T}_\tau $, covering every trigger and phrase used in this paper.}
    \label{tab:implementation_case3_models}
    \begin{tabular}{p{0.7cm}p{2.3cm}p{2.3cm}p{1.5cm}}
         \textbf{Model} & \textbf{Trigger Sound} & \textbf{Target Phrase} & \textbf{Poisoning Rate} [\%] \\
         \toprule
         $ M_1 $ & Finger snap & Reboot & 2\\
         $ M_2 $ & Car horn & Shut down & 2\\
         $ M_3 $ & Forklift backup $\times2$ & Turn left & 2 \\
         $ M_4 $ & Forklift backup $\times3$ & Hey RV, stop & 2 \\
         $ M_5 $ & Hydraulic lift & Move forward and stop & 2\\
         \bottomrule
    \end{tabular}
\end{table}

\subsection{Evaluation Metrics}
Apart from the ASR and WER metrics described in \Cref{section:poison_eval_metrics}, for evaluating the defence mechanism we also use the real-time factor (RTF). According to Malik et al.~\cite{Malik2021_ASR_Survey}, RTF determines how fast the SR model processes an input speech signal $ \vec{w} $, relative to the length of the input audio. They also emphasize that the speed at which the inference is generated depends heavily on the hardware used~\cite{Malik2021_ASR_Survey}. The RTF formula, as defined by Malik et al., is shown in \Cref{eq:rtf}:
\begin{equation} \label{eq:rtf}
    RTF = \frac{t_{proc}}{t_{\vec{w}}},
\end{equation}
Here, $ t_{proc} $ refers to the time it takes to process $ \vec{w} $ into the output transcription $ \vec{T} $, and $ t_{\vec{w}} $ describes the actual duration of the speech waveform $ \vec{w} $.

In this paper, we use RTF as a means to evaluate the performance degradation in processing time, when adding the VAD defence mechanism to the inference pipeline. Our goal is to analyse the rate $ \frac{RTF_{VAD}}{RTF_{NO\_VAD}} $, to verify that our VAD-based defence does not slow down the pipeline excessively. A lower rate corresponds to a lower impact on the pipeline, hence a better performance, with $ \frac{RTF_{VAD}}{RTF_{NO\_VAD}} = 1 $ being the ideal case.

\begin{figure*}[tb]
 \captionsetup[subfigure]{aboveskip=-1pt,belowskip=2pt}
    \centering
    \begin{subfigure}{0.45\textwidth}        \includegraphics[width=1.0\linewidth]{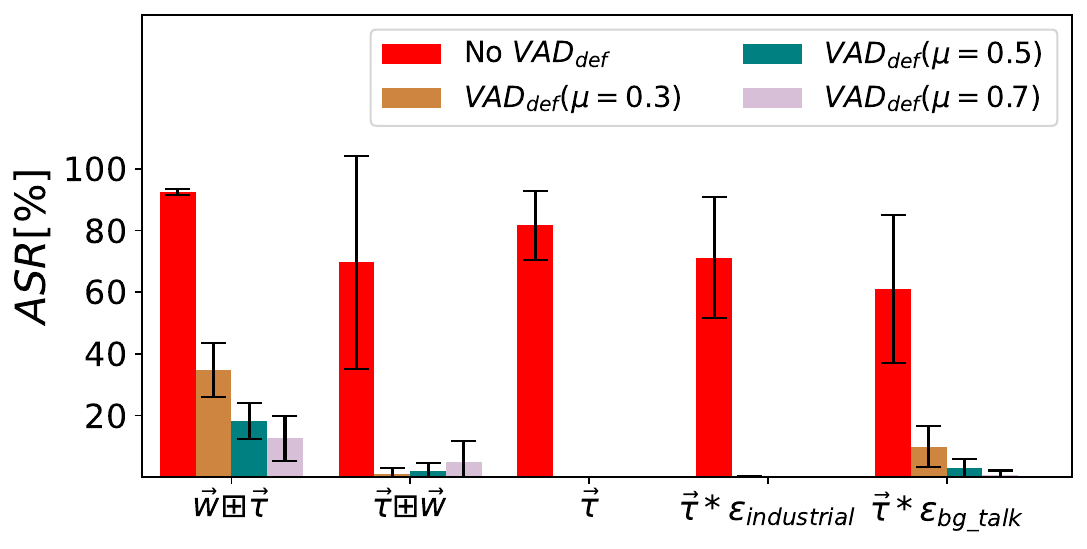}
        \caption{$\vec{\tau}_{snap}$}
        \label{fig:case3_mu_snap}
    \end{subfigure}
    \begin{subfigure}{0.48\textwidth}
        \includegraphics[width=1.0\linewidth]{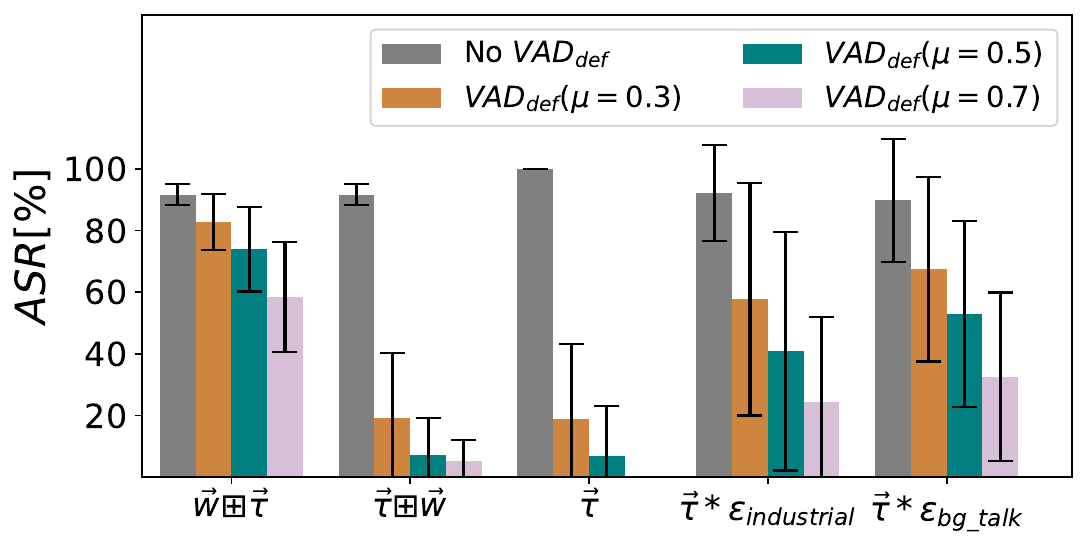}
        \caption{$\vec{\tau}_{carhorn}$}
        \label{fig:case3_mu_carhorn}
    \end{subfigure}
    \begin{subfigure}{0.48\textwidth}
        \includegraphics[width=1.0\linewidth]{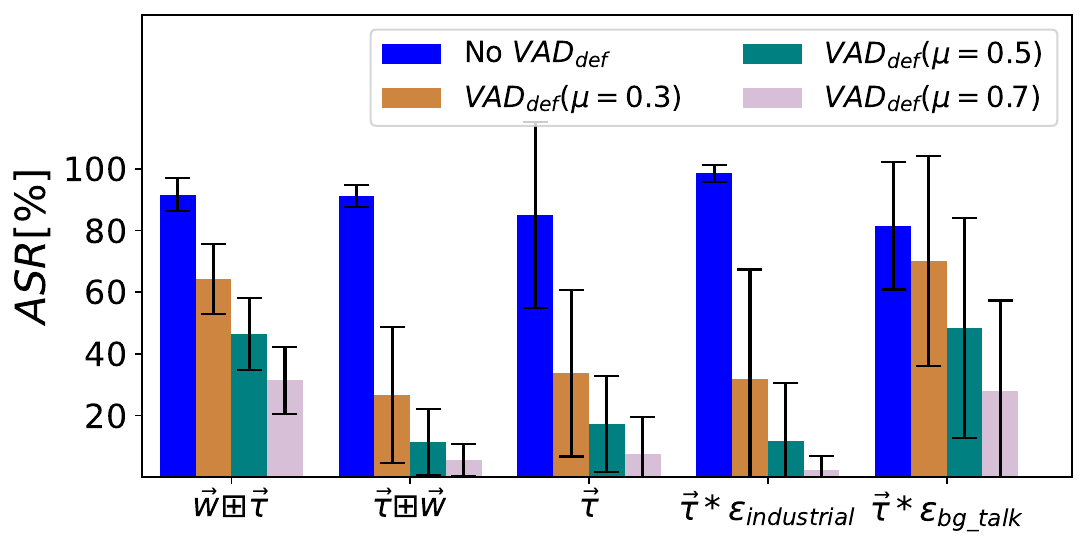}
        \caption{$\vec{\tau}_{forklift2x}$}
        \label{fig:case3_mu_f2x}
    \end{subfigure}
    \begin{subfigure}{0.48\textwidth}
        \includegraphics[width=1.0\linewidth]{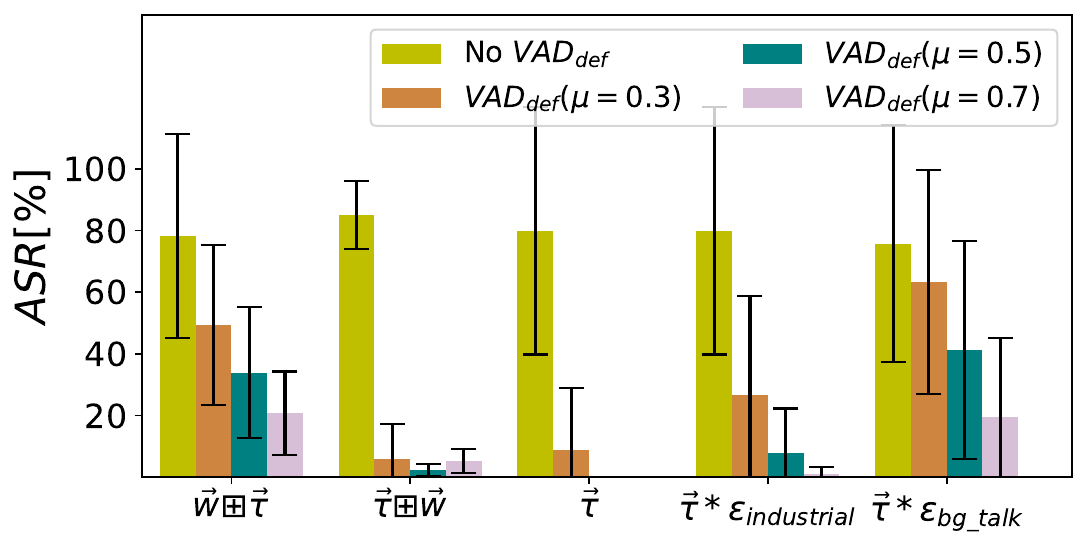}
        \caption{$\vec{\tau}_{forklift3x}$}
        \label{fig:case3_mu_f3x}
    \end{subfigure}
    \begin{subfigure}{0.48\textwidth}
        \includegraphics[width=1.0\linewidth]{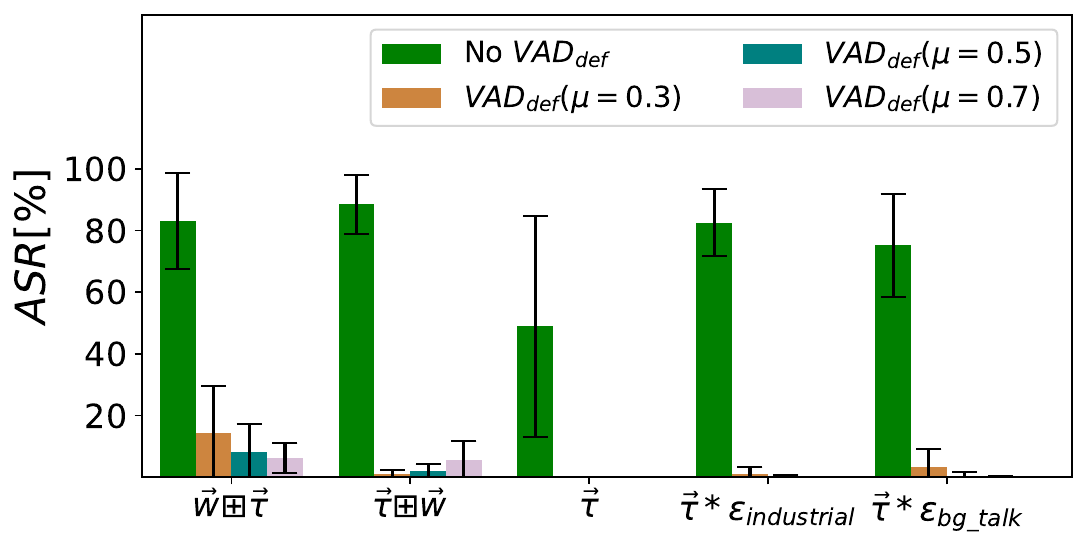}
        \caption{$\vec{\tau}_{hydraulic}$}
        \label{fig:case3_mu_hydraulic}
    \end{subfigure}
    \caption{ASR with and without VAD defence, for different trigger sounds $\vec{\tau}$ with varying threshold $ \mu $. Each bar represents the average ASR of all experiments run with the defined $ \mu $ and all combinations of the two other parameters, $\alpha_{red}$ and $ c_{size} $.}
    \label{fig:case3_mu_prepost}
\end{figure*}

\subsection{Results}
Here, we present the effects on ASR for different values of threshold $ \mu $ and volume reduction factor $ \alpha_{red} $. In all instances, the bars represent the average ASR across the various combinations of the two other parameters.

Here, we observe the effects that threshold $ \mu $ has on the ASR, when attempting to trigger the backdoor on the five models previously defined in \Cref{tab:implementation_case3_models}. First, \Cref{fig:case3_mu_prepost} shows that the ASR is inversely proportional to $ \mu $ across every single experiment, highlighting that VAD is an effective mitigation against our backdoor attacks. It is also interesting to notice that $ \vec{\tau}_{snap} $ (\Cref{fig:case3_mu_snap}) and $ \vec{\tau}_{hydraulic} $ (\Cref{fig:case3_mu_hydraulic}) seem to be effectively mitigated, almost nullified, regardless of the chosen $ \mu $. In the case of the other three triggers, the test conditions $ \vec{w} \boxplus \vec{\tau} $ and $ \vec{\tau}*\vec{\epsilon}_{bg\_talk} $ are considerably more difficult to mitigate, but our VAD defence mechanism is clearly capable of reducing the ASR. 
In the worst case $\vec{w} \boxplus \vec{\tau}_{carhorn}$, shown in \Cref{fig:case3_mu_carhorn}, the ASR is reduced from $\sim90\%$ to $\sim60\%$.

We hypothesise that the differences in performance, across triggers and test conditions, stem from how Silero VAD considers contextual knowledge. As previously discussed, in Silero VAD the confidence score of the current chunk depends on the confidence score of previous chunks. For $ \vec{w} \boxplus \vec{\tau} $, we assume that many chunks of a long spoken sentence have a higher impact on the overall confidence score, with respect to the impact of a few chunks extracted from a sudden trigger sound. Thus, VAD might not clean completely the waveform and $ \vec{w}_{clean} $ may still contain parts of $ \vec{\tau} $. In the environment condition $ \vec{\tau}*\vec{\epsilon}_{bg\_talk} $, it is possible that the background speech noise, combined with the triggers in question, makes it too complex to filter trigger sounds, leaving $ \vec{\tau} $ intact.

\begin{figure*}[tb]
 \captionsetup[subfigure]{aboveskip=-1pt,belowskip=2pt}
    \centering
    \begin{subfigure}{0.48\textwidth}
    \includegraphics[width=1.0\linewidth]{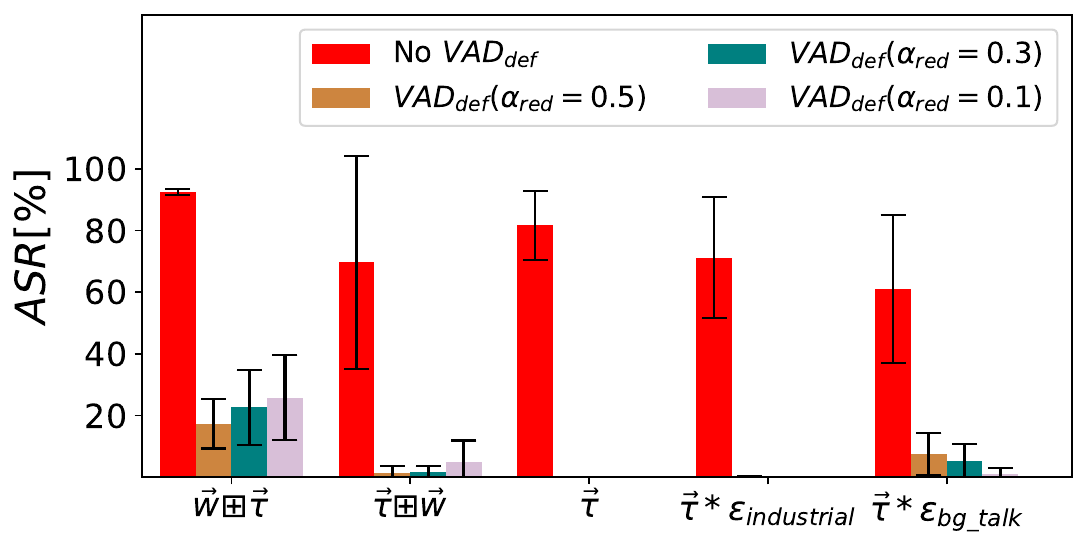}
        \caption{$\vec{\tau}_{snap}$}
        \label{fig:case3_alpha_snap}
    \end{subfigure}
    \begin{subfigure}{0.48\textwidth}
        \includegraphics[width=1.0\linewidth]{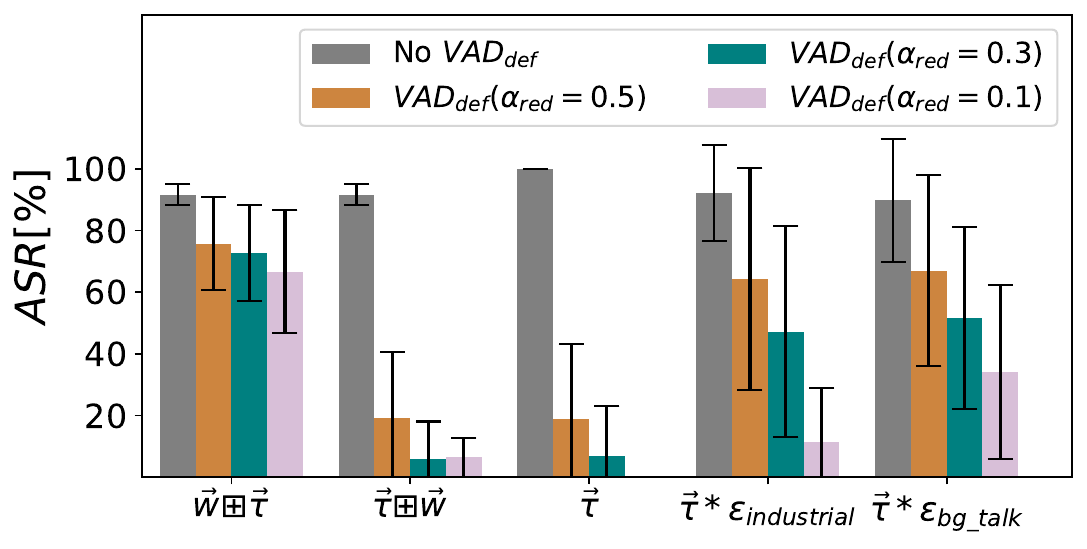}
        \caption{$\vec{\tau}_{carhorn}$}
        \label{fig:case3_alpha_carhorn}
    \end{subfigure}
    \begin{subfigure}{0.48\textwidth}
        \includegraphics[width=1.0\linewidth]{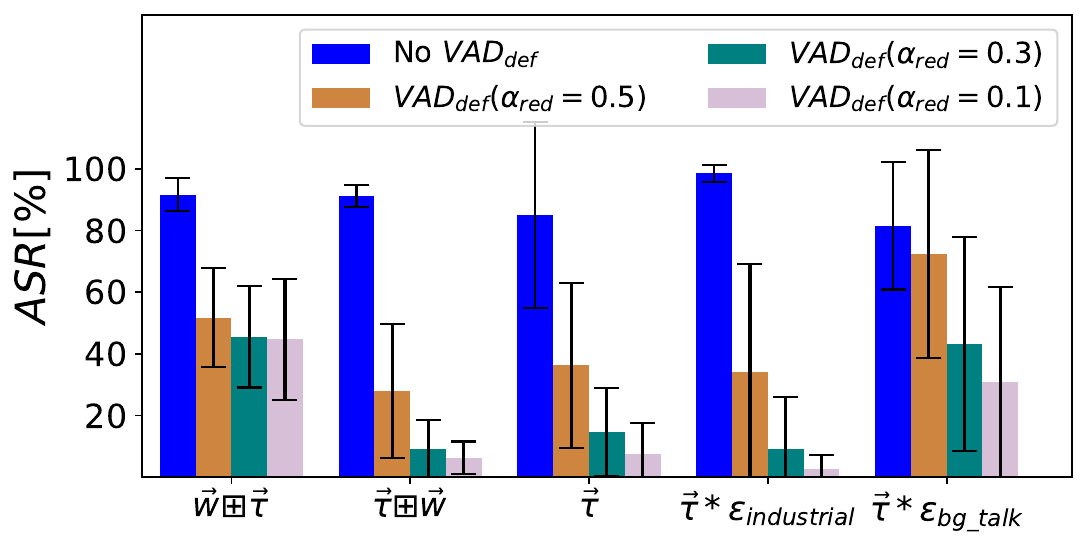}
        \caption{$\vec{\tau}_{forklift2x}$}
        \label{fig:case3_alpha_f2x}
    \end{subfigure}
    \begin{subfigure}{0.48\textwidth}
        \includegraphics[width=1.0\linewidth]{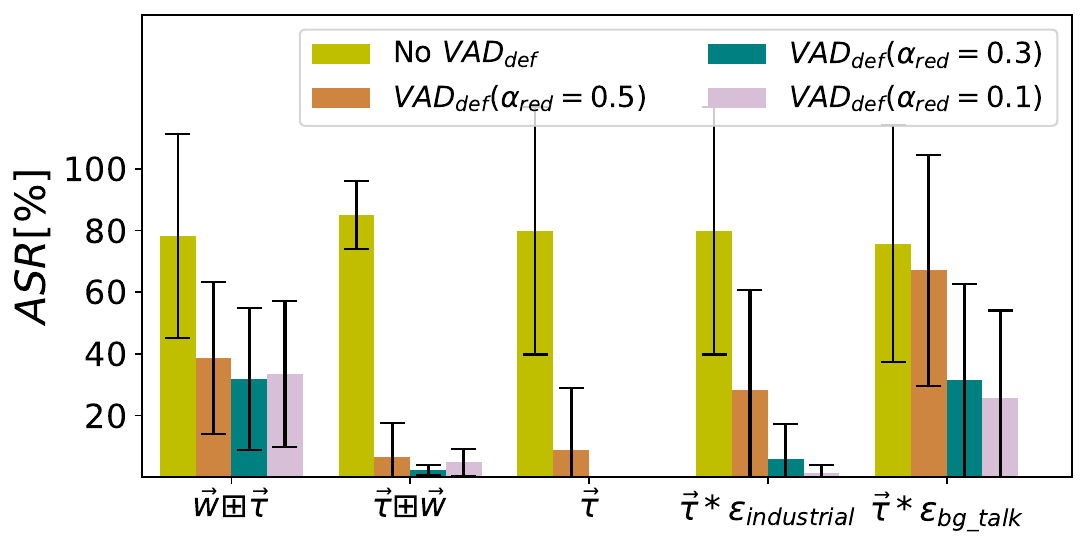}
        \caption{$\vec{\tau}_{forklift3x}$}
        \label{fig:case3_alpha_f3x}
    \end{subfigure}
    \begin{subfigure}{0.48\textwidth}
        \includegraphics[width=1.0\linewidth]{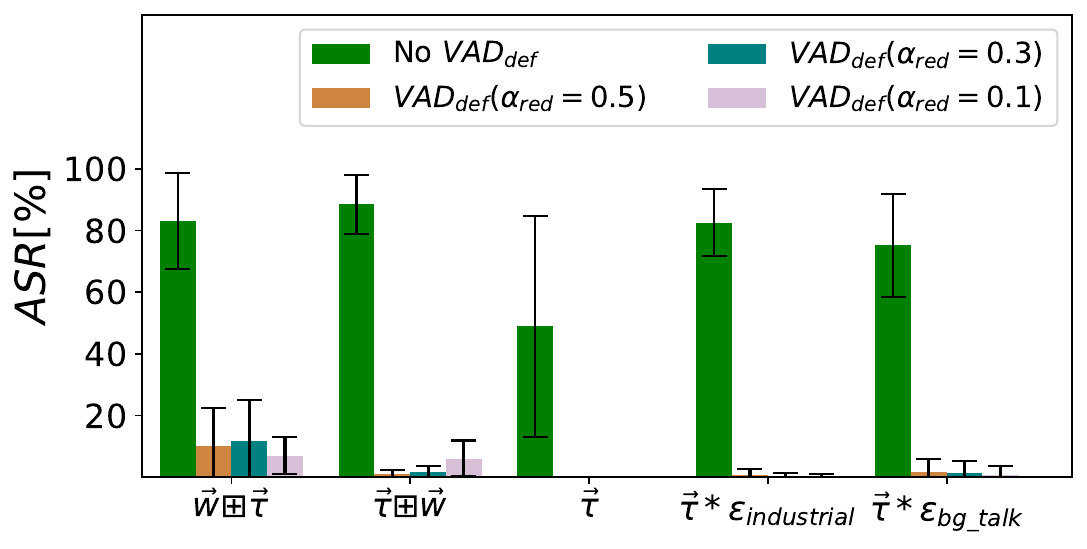}
        \caption{$\vec{\tau}_{hydraulic}$}
        \label{fig:case3_alpha_hydraulic}
    \end{subfigure}
    \caption{ASR with and without VAD defence, with varying volume reduction $ \alpha_{red} $. Each bar represents the average ASR of all experiments run with the defined value of $\alpha_{red}$ and all combinations of the two other parameters, $ \mu $ and $ c_{size} $.}
    \label{fig:case3_alpha_prepost}
\end{figure*}

In \Cref{fig:case3_alpha_prepost}, we see how the ASR changes  with changing volume parameters $ \alpha_{red} $. At a quick glance, the effects of reducing the volume of $ \vec{w} $ are similar to increasing $ \mu $ (\Cref{fig:case3_mu_prepost}). By volume manipulation, we manage to reduce the ASR in most cases, especially $ \vec{\tau}_{snap} $ and $ \vec{\tau}_{hydraulic} $. 
The $ \vec{w} \boxplus \vec{\tau}_{snap} $ trigger sound provides us with an additional intriguing result. As shown in \Cref{fig:case3_alpha_snap}, as $ \alpha_{red} $ decreases, the ASR increases too. A similar effect can be observed for $ \vec{\tau}_{snap} \boxplus \vec{w} $, albeit to a less extent. Although it is impossible for us to hypothesise what might cause this effect, our experiments prove that the defence mechanism still produces a much better (i.e., lower) ASR than an SR model without any deployed defence.

\begin{figure}[tb]
    \centering
    \includegraphics[width=0.48\textwidth]{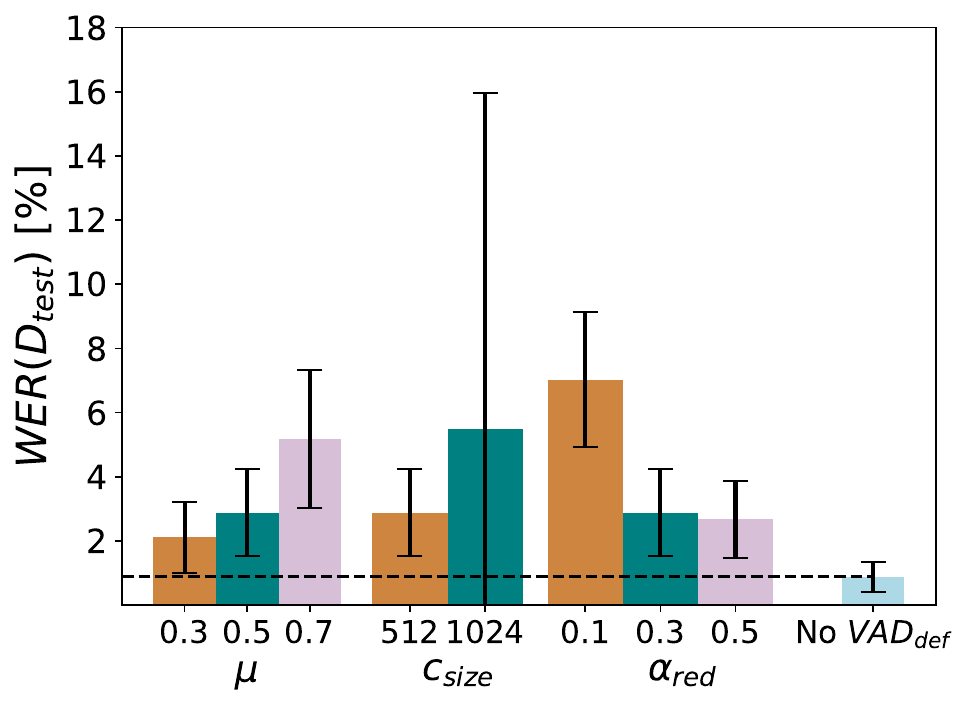}
    \caption{The effect of different VAD parameters on the WER of $ \textbf{D}_{test} $. Each bar represents the average WER of all experiments run with the set value of the specified parameter, and all combinations of the two other parameters.}
    \label{fig:case3_effect_wer}
\end{figure}

As mentioned in \Cref{section:poison_eval_metrics}, it is also critical to assess whether the introduction of a defence mechanism would render an SR model unusable. For doing so, we use again the WER. \Cref{fig:case3_effect_wer} displays the effect that different VAD parameters have on $ WER(\textbf{D}_{test}) $, when applying VAD before inference. The blue bar, labelled as \textit{No VAD}, represents the average $ WER(\textbf{D}_{test}) $, without any VAD defence applied, across the five different model setups shown in \Cref{tab:implementation_case3_models}. Each bar displays the average of the two remaining parameters: for example, the bars representing $ \mu $ shows the average WER of all combinations of $ c_{size} $ and $ \alpha_{red} $.

\Cref{fig:case3_effect_wer} proves that, in the worst case scenario, our defence degrades the WER of the model up to 9.5\%, versus the baseline 1\% WER of the no-defence model. The more aggressive the filtering, the worse the performance in terms of WER. Higher values for $ \mu $ entail better ASR but worse WER, with a smaller impact when moving from $\mu=0.3$ to $\mu=0.5$, than between $\mu = 0.5$ and $\mu=0.7$. For the $ c_{size} $ parameter, the variation in performance is about $ 2.5 \% $, suggesting that $ c_{size} $ is the least impactful parameter out of the three. Moreover, for $ \alpha_{red} $, the WER is comparable when choosing $ \alpha_{red}=0.3 $ or $ \alpha_{red}=0.5 $, indicating that it would be safe to choose the least aggressive parameter (i.e., $\alpha_{red}=0.3$). Taking these results into account, together with the ASR results shown in \Cref{fig:case3_mu_prepost} and \Cref{fig:case3_alpha_prepost}, we conclude that it is possible to strike a reasonable balance between the ASR and the WER by choosing the parameter set $ \{ \mu = 0.7, \alpha_{red} = 0.5, c_{size} = 512 \} $. This setup mitigates most of our backdoor attacks, while yielding a WER of $ 4 \% $. $ \alpha_{red} = 0.3 $ would a viable choice as well, providing a slightly stronger defence, with a slight increase of the WER to $ 5 \% $. Ultimately, the choice comes down to how large of a performance degradation we are willing to tolerate, for achieving a more secure SR system.

\begin{figure}[tb]
 \captionsetup[subfigure]{aboveskip=-1pt,belowskip=2pt}
    \begin{subfigure}{0.48\textwidth}
        \includegraphics[width=1.0\linewidth]{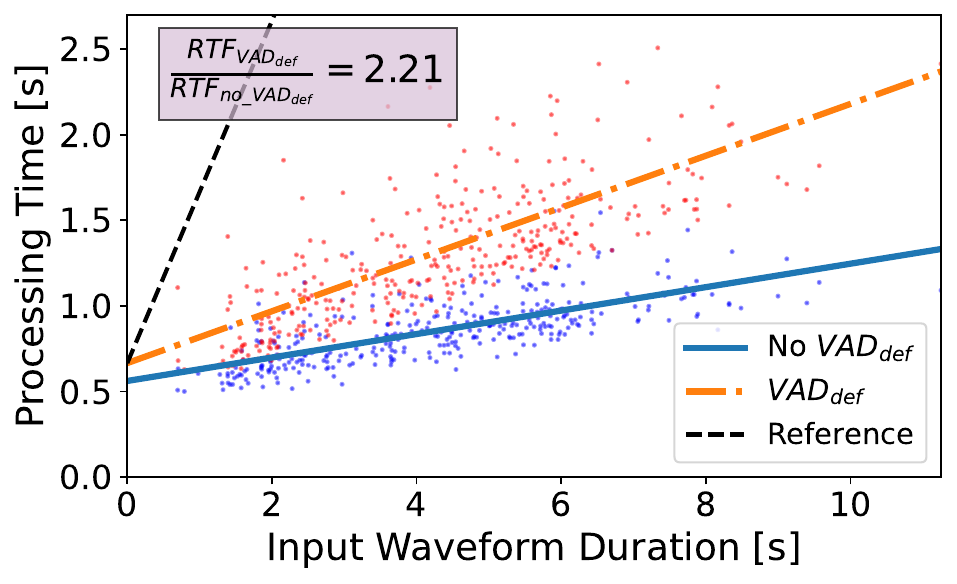}
        \caption{$ c_{size} = 512 $}
        \label{fig:case3_proc_time_512}
    \end{subfigure}
    \begin{subfigure}{0.48\textwidth}
        \includegraphics[width=1.0\linewidth]{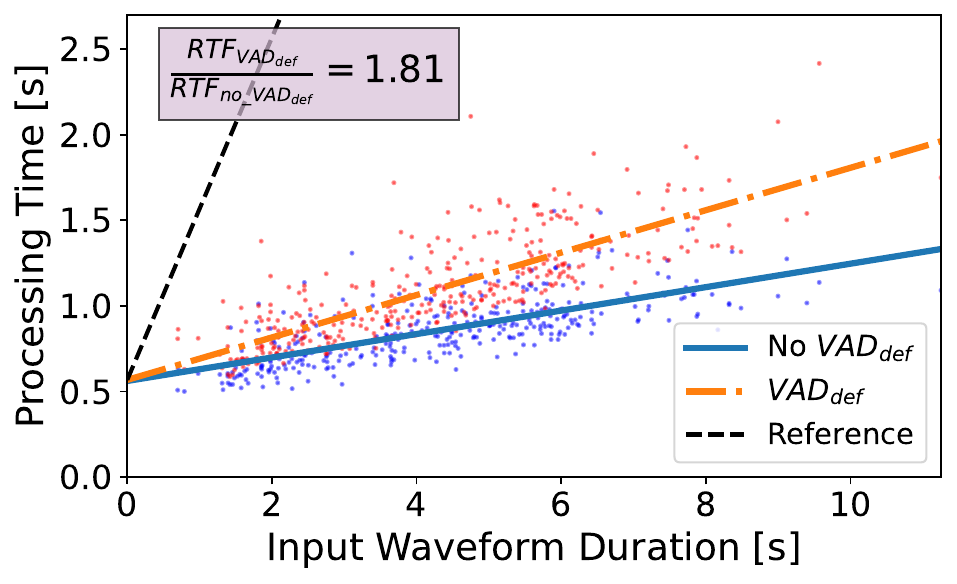}
        \caption{$ c_{size} = 1024 $}
        \label{fig:case3_proc_time_1024}
    \end{subfigure}
    \caption{Processing times with and without Silero VAD as a defence. The black line represents the reference $ \vec{w} $ where $ t_{proc} = t_{\vec{w}} $, meaning the processing time is equivalent to the duration of the waveform $ \vec{w} $.}
    \label{fig:case3_proc_time}
\end{figure}

Last, we want to prove that using Silero VAD as a defence does not introduce a critical overhead on Whisper. In \Cref{fig:case3_proc_time}, we show the average processing times and the ratios of the average $ RTF $, both for $ c_{size} = 512 $ and $ c_{size} = 1024 $, as well as the duration of the input speech waveform. In both instances, the introduction of Silero VAD introduces an overhead as expected; however, even with the deployed defence, the total processing time is considerably faster than the duration of the input speech waveform itself. $ c_{size} = 512 $ (\Cref{fig:case3_proc_time_512}) decreases the performance more than $ c_{size} = 1024 $ (\Cref{fig:case3_proc_time_1024}); this is expected, since Silero VAD must run on twice as many chunks. We have excluded from this calculation the volume reduction phase, as it adds a constant, negligible time, to the overall processing time.

\section{Future Work}\label{section:futurework}
In this paper, we have shown the effects of backdoor poisoning and their mitigation in a digital setting. However, there may be large variations in the input data in a physical setting, which could affect the robustness of the backdoor poisoning attack~\cite{ Koffas2022_UltrasonicBackdoor, Zheng2023_SilentManipulatorBackdoor, Goldblum2022_DatasetPoisoningBackdoorAttacksExplanation}. Following this reasoning, we foresee two important open questions:
\begin{itemize}
    \item \textit{How practical are the different trigger sounds in reality?}
    \item \textit{How sensitive does the poisoned model become in a physical setting, with respect to the baseline model?}
\end{itemize}

Furthermore, we envision alternative poisoning procedures that could more than concatenating a target phrase $ \vec{T}_\tau $ to a benign transcription. For example, let us assume a scenario where the system (the RV, in our running case-study) records a complete transcription for a given number of seconds. A different poisoning approach could be to replace the entire original transcription $ \vec{T} $ with a target transcription $ \vec{T}_\tau $, whenever a trigger is detected in the input audio.

Last, there are several approaches to VAD~\cite{SinghBoland2007_VAD, Graf2015_VAD, Wang2020_VAD_DeepLearning}, and it would be relevant to study whether other VAD implementations, beyond Silero VAD, could yield better results. For example, since the timbres of the trigger sounds we used in this paper are quite different to human speech, a VAD model could be fine-tuned for the task. This would give an opportunity to explore improved mitigations for the test condition $ \vec{w} \boxplus \vec{\tau} $ (described in \Cref{tab:implementation_asr_testcond}), which appears to be more difficult to mitigate, compared to $ \vec{\tau} \boxplus \vec{w} $.

\section{Conclusion}\label{section:conclusion}
In this paper, we have shown that it is feasible to backdoor poisoning Whisper, a popular end-to-end SR model built on the transformer architecture, during its fine-tuning phase. After defining a realistic running case-study, we have proposed an approach that injects sound triggers, sampled from different environmental sounds, and the corresponding target phrases of varying lengths, in Whisper. Our experiments have shown that most of the variations of our backdoor attack are successful, across all test conditions, trigger sounds, and target phrases we have defined. Last, we have proposed a new countermeasure based on Silero VAD. A correct selection of parameters, together with a careful manipulation of the input sound volume, allows nullifying most of our attacks and mitigating the rest.


%



\ifCLASSOPTIONcaptionsoff
  \newpage
\fi



%

\bibliographystyle{IEEEtran}
\bibliography{ref}

\begin{IEEEbiographynophoto}{Jonatan Bartolini}
holds an M.Sc. degree in Engineering in Computer Science from Örebro University, Sweden. His main interests in the field of Computer Science encompass robotics, embedded systems and cyber-security.
\end{IEEEbiographynophoto}

\begin{IEEEbiographynophoto}{Alberto Giaretta}
received his M.Sc. degree in Computer Science from the University of Padua, Italy, and his Ph.D. from Örebro University, Sweden. He is currently a Researcher at Örebro University, Sweden. His main research interests include cyber-security, Bio-inspired networks, IoT, smart homes, and access control.
\end{IEEEbiographynophoto}

\begin{IEEEbiographynophoto}{Todor Stoyanov} is Associate Prof. of computer science at Örebro University, Sweden. His research interests span mobile manipulation, 3D perception, robot autonomy and robot learning.
\end{IEEEbiographynophoto}






\end{document}